\documentclass[12pt]{JHEP3}
\usepackage{epsfig}
\preprint{ {\tt hep-th/0602254} }

\newcommand{\be}[1]{ \begin{equation}\label{#1} }
\newcommand{\ee}{\end{equation}}
\newcommand{\bea}[1]{\begin{eqnarray}\label{#1} }
\newcommand{\eea}{\end{eqnarray}}
\newcommand{\eq}[1]{(\ref{#1})}

\newcommand{\II}{{\cal I}}
\newcommand{\JJ}{{\cal J}}
\newcommand{\AAA}{{\cal A}}
\newcommand{\III}{{\cal I}}
\newcommand{\FF}{{\cal F}}
\newcommand{\NN}{{\cal N}}
\newcommand{\p}{\partial}
\newcommand{\wt}{\widetilde}
\def\ZZZ{{\hskip-3pt\hbox{ Z\kern-1.6mm Z}}}
\def\zzz{{\hskip-3pt\hbox{ z\kern-1mm z}}}

\title{Product Representation of Dyon Partition Function in CHL Models}
\author{Justin R. David, Dileep P. Jatkar and Ashoke Sen \\
Harish-Chandra Research Institute, \\
Chhatnag Road., Jhunsi, \\
Allahabad 211019, India.
\\
\email{justin,dileep,sen@mri.ernet.in}
}

\abstract{ A formula for the exact partition function of 1/4 BPS dyons
  in a class of CHL models has been proposed earlier. The formula
  involves inverse of Siegel modular forms of subgroups of
  $Sp(2,\ZZZ)$. In this paper we propose product formulae for these
  modular forms. This generalizes the result of Borcherds
  and Gritsenko and Nikulin
  for the weight 10 cusp form of the full $Sp(2,\ZZZ)$ group.}

\begin{document}
 
\baselineskip 4ex

\section{Introduction and Summary} \label{s0}

There exists a proposal for the exact degeneracy of dyons in
toroidally compactified heterotic string
theory\cite{9607026,0412287,0505094,0506249,0508174} and also in
toroidally compactified type II string theory\cite{0506151}. These
formul\ae\ are invariant under the S-duality transformations of the
theory, and also reproduce the entropy of a dyonic black hole in the
limit of large charges\cite{0412287}.  In \cite{0510147} this proposal
was generalized to a class of CHL
models\cite{CHL,CP,9507027,9507050,9508144,9508154}, obtained by
modding out heterotic string theory on $T^2\times T^4$ by a $\ZZZ_N$
transformation that involves $1/N$ unit of translation along one of
the circles of $T^2$ and a non-trivial action on the internal
conformal field theory (CFT) describing heterotic string
compactification on $T^4$. The values of $N$ considered in
\cite{0510147} were $N=$2,3,5,7.  Using string-string
duality\cite{9410167,9501030,9503124,9504027,9504047} one can relate
these models to $\ZZZ_N$ orbifolds of type IIA string theory on
$T^2\times K3$, with the $\ZZZ_N$ transformation acting as $1/N$ unit
of shift along a circle of $T^2$ together with an action on the
internal CFT describing type IIA string compactification on $K3$.

The proposal of \cite{0510147} may be summarized as follows.
If we denote by $Q_e$ and $Q_m$ the electric and the magnetic
charge vectors then the degeneracy $d(Q_e, Q_m)$ of dyons
carrying charges $(Q_e,Q_m)$ is of the form
\be{e01}
d(Q_e,Q_m) =  g\left({1\over 2}Q_m^2, {1\over 2}
     Q_e^2, Q_e\cdot Q_m\right)\, ,
\ee
where $g(m,n,p)$ is defined through the Fourier expansion
\be{ei44.2}
   {1\over \wt\Phi_k(U,T,V)} =C_0\, 
   \sum_{m,n,p\atop m\ge -1,n\ge -1/N}
   e^{2\pi i(mU
     + nT + pV)} g(m,n,p)\, .
\ee
Here $C_0$ 
is a numerical constant and $\wt\Phi_k(U,T,V)$ is a
modular form of weight $k$ under a subgroup
$\wt G$ of $Sp(2,\ZZZ)\equiv SO(2,3;\ZZZ)$ 
where
\be{es01.5}
k = {24\over N+1}-2\, .
\ee
An explicit algorithm for constructing the Fourier expansion of
$\wt\Phi_k$ in the variables $T$, $U$ and $V$ was given in
\cite{0510147}.

The degeneracy $d(Q_e,Q_m)$ defined through eqs.(\ref{e01}),
(\ref{ei44.2}) is invariant under the T- and S-duality symmetries of
the theory.  Furthermore it generates integer results for the
degeneracies and its behaviour for large charges is consistent with
the black hole entropy calculation\cite{0510147,0601108}.

In this paper we use the method of \cite{9512046,9607029} to propose
an alternative form of $\wt\Phi_k$ as an infinite product:
\bea{es05}
 \wt\Phi_{k}(U,T,V) &=& -(i\sqrt N)^{-k-2}\, 
\, \exp\left(2\pi i \left( {1\over N}\,
T+ U + V\right) \right) \nonumber \\
&&
\prod_{r=0}^{N-1}
\prod_{l,b\in \zzz, k'\in \zzz+{r\over N}\atop
k',l,b>0}\Bigg\{ 
 1 - \exp( 2\pi i( k'T + l U + bV)) \Bigg\}^{{1\over 2}\sum_{s=0}^{N-1}
e^{-2\pi ils/N}\, c^{(r,s)}(4lk' -b^2)
  }  \nonumber \\
  && \prod_{r=0}^{N-1}
\prod_{l,b\in \zzz, k'\in \zzz-{r\over N}\atop
k',l,b>0}\Bigg\{ 
 1 - \exp( 2\pi i( k'T + l U + bV)) \Bigg\}^{{1\over 2}\sum_{s=0}^{N-1}
e^{2\pi ils/N}\, c^{(r,s)}(4lk' -b^2)
  }  \nonumber \\
 \eea
where $(k', l , b)>0$ means $k'> 0 , l\geq 0, b \in\ZZZ$ or $ k' =0,
l>0, b \in\ZZZ$ or $k' =0, l =0, b <0$ and $c^{(r,s)}(n)$ are some
calculable coefficients related to the twisted elliptic genus of $K3$.
If $\wt g$ denotes the generator of the $\ZZZ_N$ action on $K3$ that
is used in the construction of the CHL model, then we define the
twisted elliptic genus of $K3$ as
\be{es06}
F^{(r,s)}(\tau,z) = {1\over N} \, Tr^{K3}_{RR; \wt g^r}
\left( (-1)^{F_{K3}}
(-1)^{\bar F_{K3}} \, \wt g^s\, 
e^{2\pi i z F_{K3}} \, q^{L_0} \bar q^{\bar L_0}
\right)\, , \quad 0\le r,s\le (N-1)\, ,
\ee
where $Tr^{K3}_{RR; \wt g^r}$ denotes trace in the superconformal
field theory associated with target space $K3$ in the $\wt g^r$
twisted RR sector, $q=e^{2\pi i \tau}$, and $F_{K3}$, $\bar F_{K3}$
denote the left- and right-handed world-sheet fermion numbers in this
theory.  Here and throughout the rest of the paper $L_0$ and $\bar
L_0$ include an additive factor of $-c/24$ so that the RR sector
ground state has $L_0=\bar L_0=0$.  The coefficients $c^{(r,s)}(n)$
are then defined through the Fourier expansion of $F^{(r,s)}(\tau,z)$:
\be{es07}
F^{(r,s)}(\tau,z) =\sum_{b\in \zzz, n }
c^{(r,s)}(4n-b^2)\,  q^{n} \, 
e^{2\pi i z b}\, .
\ee
Furthermore for the $N=2$, $k=6$ case we were able to explicitly compute 
the functions $F^{(r,s)}(\tau,z)$. They are given by
\bea{ex4pr}
&& F^{(0,0)}(\tau, z) = 4\left[ {\vartheta_2(\tau,z)^2
\over \vartheta_2(\tau,0)^2} +
{\vartheta_3(\tau,z)^2\over \vartheta_3(\tau,0)^2}
+ {\vartheta_4(\tau,z)^2\over \vartheta_4(\tau,0)^2}\right]\, ,
\nonumber \\
&& F^{(0,1)}(\tau, z) =  4 \, {\vartheta_2(\tau,z)^2\over
\vartheta_2(\tau,0)^2} \, , \quad 
F^{(1,0)}(\tau, z) = 4 {\vartheta_4(\tau,z)^2\over \vartheta_4(\tau,0)^2}
\, ,
\quad 
F^{(1,1)}(\tau, z) =  4\, {\vartheta_3(\tau,z)^2\over
\vartheta_3(\tau,0)^2} \, . \nonumber \\
\eea
For higher values of $N$ we did not evaluate the functions
$F^{(r,s)}(\tau,z)$ directly, but were able to guess their
forms from general considerations. The results are:
\bea{fifthn}
F^{(0,0)}(\tau, z) &=& {8\over N} A(\tau, z)\, ,
\nonumber \\
F^{(0,s)}(\tau, z) &=& {8\over N(N+1)} \, A(\tau, z) -{2\over N+1}
\, B(\tau, z) \, E_N(\tau) \qquad \hbox{for $1\le s\le (N-1)$}
\, , \nonumber \\
F^{(r,rk)}(\tau, z) &=& {8\over N(N+1)} \, A(\tau, z)
+ {2\over N(N+1)} \, E_N\left({\tau+k\over N}\right)\, B(\tau, z)\, ,
\nonumber \\
&& \qquad \qquad \qquad \qquad 
\qquad \hbox{for $1\le r \le (N-1)$, $0\le k\le (N-1)$}
\, ,\nonumber \\
\eea
where
\be{efirstn}
 A(\tau, z) =  \left[ {\vartheta_2(\tau,z)^2
\over \vartheta_2(\tau,0)^2} +
{\vartheta_3(\tau,z)^2\over \vartheta_3(\tau,0)^2}
+ {\vartheta_4(\tau,z)^2\over \vartheta_4(\tau,0)^2}\right]\, ,
\ee
\be{secondn}
B(\tau, z) = \eta(\tau)^{-6} \vartheta_1(\tau, z)^2\, ,
\ee
and
\be{thirdn}
E_N(\tau) = {12 i\over \pi(N-1)} \, \p_\tau \left[ \ln\eta(\tau)
-\ln\eta(N\tau)\right]= 1 + {24\over N-1} \, \sum_{n_1,n_2\ge 1\atop
n_1 \ne 0 \,  mod \, N} n_1 e^{2\pi i n_1 n_2 \tau}\, .
\ee

Eq.(\ref{es05}) gives a generalization of Borcherds
and Gritsenko and Nikulin's
result\cite{borcherds,9504006} of the 
product representation of
$\wt\Phi_{10}$, -- the unique cusp form of weight 10 of the group
$Sp(2,\ZZZ)$.  A systematic procedure for arriving at the product
representation for $\wt\Phi_{10}$ was given in \cite{9512046}.  Our
construction of $\wt\Phi_k$ is essentially based on a generalization
of the techniques of \cite{9512046}.

Given the two different constructions of $\wt\Phi_k$, -- one given in
\cite{0510147} and one in the present paper, it is natural to ask if
they are the same. For the $N=2$, $k=6$ case we have compared 31
different Fourier expansion coefficients of the two proposals and
found them to be the same.\footnote{Actually we compare not the
  Fourier expansion coefficients of $\wt\Phi_k$ but those of a closely
  related object $\Phi_k(U,T,V)=T^{-k} \, \wt\Phi_k(U - T^{-1}V^2,
  -T^{-1}, T^{-1}V)$. } \label{f1} For other values of $N$ we have
compared the expansions up to order $e^{4\pi i T} e^{4\pi i U}$ and
all powers of $e^{2\pi i V}$.  For general $N$ we also verify that the
behaviour of $\wt\Phi_k$ (and of $\Phi_k$ introduced in footnote
\ref{f1}) in the $V\to 0$ limit as well as in the
$U\to i\infty$ limit agrees with the results found in
\cite{0510147}.

The rest of the paper is organized as follows. In section \ref{s1} we
outline the strategy that we shall be using for finding $\wt\Phi_k$.
Sections \ref{s2} and \ref{s3} involve detailed calculations leading
to the determination of $\wt\Phi_6$ associated with the $\ZZZ_2$
orbifold theory. In section \ref{s3.5} we give the final form of
$\wt\Phi_6$ and compare some of its properties with those found in
\cite{0510147}.  Section \ref{s4} is devoted to the construction of
the related quantity $\Phi_6$ described in footnote \ref{f1} and its
comparison with the corresponding quantity calculated in
\cite{0510147}.  In section \ref{s5} we describe the construction of
$\wt\Phi_k$ and $\Phi_k$ for a general $k$ given in (\ref{es01.5}).
The three appendices contain some technical details which were omitted
from discussion in the main body of the paper.

\section{The Strategy} \label{s1}

Our goal is to find a product representation for $\wt\Phi_k$.  In
attaining this goal we shall proceed as in the case of ordinary
toroidal compactification of heterotic string theory or equivalently
type II string theory on $T^2\times K3$. This corresponds to the case
$N=1$, $k=10$ and the associated modular form $\wt\Phi_{10}$ is the
unique weight 10 cusp form of the Siegel modular group $Sp(2;\ZZZ)$.
The steps leading to a systematic construction of the product
representation of $\wt\Phi_{10}$ are as follows\cite{9512046}:

\begin{enumerate}
  
\item We consider a superconformal $\sigma$-model with target space
  $T^2\times K3$ with $y^1, y^2$ denoting the $T^2$ coordinates.  We
  denote by $F_{K3}$ and $F_{T^2}$ the holomorphic parts of the
  world-sheet fermion number associated with the $K3$ and the $T^2$
  parts and by $\bar F_{K3}$ and $\bar F_{T^2}$ the anti-holomorphic
  parts of the world-sheet fermion number associated with the $K3$ and
  the $T^2$ parts.  We shall be considering an arbitrary $T^2$
  parametrized by the K\"{a}hler modulus $T$ and complex structure modulus
  $U$, and arbitrary Wilson lines $A_1$, $A_2$ corresponding to
  deforming the world-sheet theory by the marginal operator
\be{emarginal}
\sum_{i=1}^2 A_i \int d^2 z \, 
\bar\p Y^i J_{K3}\, ,
\ee
where $J_{K3}$ is the U(1) current corresponding to the charge
$F_{K3}$.  We shall denote by $V$ the complex combination $A_2-iA_1$.
$V$ is normalized so that $V\equiv V+1$.

This theory has an $SO(2,3;\ZZZ)$ T-duality group. If we denote by
$(m_1, m_2)$ the integers labeling momenta along $y^1,y^2$, by
$(n_1, n_2)$ the integers labeling winding along $y^1, y^2$, and
by $b$ the $F_{K3}$ charge, then the $SO(2,3;\ZZZ)$ transformation
$S$ acts on these charges and the parameters $T$, $U$, $V$ as
\be{eso32}
\pmatrix{m_1'\cr m_2'\cr n_1'\cr n_2'\cr b'}
= S \pmatrix{m_1\cr m_2\cr n_1\cr n_2\cr b}\, ,
\qquad \pmatrix{T'\cr T'U'-V^{\prime2}\cr -U'
\cr 1\cr 2V'}=
\lambda \, S\, \pmatrix{T\cr TU-V^2\cr -U\cr 1\cr 2V}\, 
\ee
where $S$ is a $5\times 5$ matrix with integer entries, satisfying
\be{eso33}
S^T L S=L, \qquad L = \pmatrix{0 & I_2 & 0\cr I_2 & 0 & 0\cr
0 & 0 & 1/2}\, ,
\ee
and $\lambda$ is a number to be adjusted so that the fourth element of
the vector on the right hand side of (\ref{eso32}) is 1. $I_n$ denotes
$n\times n$ identity matrix.

Using the equivalence between $SO(2,3)$ and $Sp(2)$ we can represent
the T-duality group elements by $Sp(2,\ZZZ)$ matrices of the form
$\pmatrix{A & B\cr C & D}$ where $A$, $B$, $C$ and $D$ are each
$2\times 2$ matrix with integer entries satisfying
\be{ek3}
 AB^T=BA^T, \qquad  CD^T=DC^T, \qquad 
 AD^T-BC^T=I_2\, .
\ee 
If we define
\be{ef5}
\Omega=\pmatrix{U & V \cr V & T}
\, ,
\ee
then the duality group acts on $\Omega$ as
\be{ek4}
\Omega\to (A\Omega+B)(C\Omega+D)^{-1}\, .
\ee

\item In this theory we define:
\be{ef1old}
\II_0(U,T,V) = \int_{\FF} {d^2\tau\over \tau_2} \, Tr_{RR}\left(
(-1)^{(F_{K3} +
F_{T^2})}
(-1)^{(\bar F_{K3}+\bar F_{T^2})} F_{T^2} \bar F_{T^2}
q^{L_0} \bar q^{\bar L_0} 
\right)
\ee
where $\FF$ is the fundamental domain of $SL(2,\ZZZ)$ and $q=e^{2\pi i
  \tau}$.  $\II(U,T,V)$ is expected to be invariant under
$SO(2,3;\ZZZ)$ transformation.

\item Analysis of the integral given in (\ref{ef1old}) shows that it can
be expressed in the form
\be{ek5}
\II_0 = -20\ln
 \det\hbox{Im}\,\Omega - 2\ln\wt\Phi_{10}(\Omega) 
 - 2\ln\wt\Phi_{10} 
(\bar\Omega) \, + \, \hbox{constant}
\ee
where $\wt\Phi_{10}(\Omega)$ is a 
holomorphic function of $T$, $U$
and $V$ with a product representation. Since under the duality
transformation (\ref{ek4})
\be{ek6}
\det \hbox{Im}\, \Omega \to (\det (C\Omega+D))^{-1}\, 
(\det (C\bar\Omega+D))^{-1}\, \det \hbox{Im}\, 
\Omega\, , 
\ee
and $\II_0$ is invariant,
we must have\footnote{In principle there could be $\Omega$
independent phases on the right hand side of (\ref{ek7}), but it
is known that they are absent in this case.}
\be{ek7}
\wt\Phi_{10}\left((A\Omega+B)(C\Omega+D)^{-1}\right)
= (\det (C\Omega+D))^{10}\, \wt\Phi_{10}(\Omega)\, .
\ee
Thus $\wt\Phi_{10}(\Omega)$ must be a Siegel modular form of
weight 10. This leads to the construction of the product representation
of $\wt\Phi_{10}$.

\end{enumerate}

Our goal is to construct a modular form $\wt\Phi_k$ of weight $k$ of
an appropriate subgroup $\wt G$ of $SO(2,3;\ZZZ)$ for $k$ given in
(\ref{es01.5}). The subgroup $\wt G$ is the T-duality group of the
superconformal field theory with target space $(T^2\times K3)/\ZZZ_N$
where the $\ZZZ_N$ acts as a $1/N$ unit of shift along a circle on
$T^2$ and as a geometric transformation of order $N$ on
$K3$.\footnote{In order to preserve the $\NN=4$ target space
  supersymmetry, the $\ZZZ_N$ action on $K3$ must commute with the
  (4,4) superconformal symmetry possessed by a supersymmetric
  $\sigma$-model with target space $K3$.}  Thus only those
$SO(2,3;\ZZZ)$ transformation which commute with the $1/N$ unit of
shift along $T^2$ will be symmetries of the resulting theory.

We shall try to construct $\wt\Phi_k$ by first defining an analog of
the integral $\II_0$ invariant under this subgroup and then splitting
it into a sum of an holomorphic piece, an anti-holomorphic piece and a
term proportional to $\ln\det\hbox{Im}\,\Omega$ as in (\ref{ek5}).  A
natural candidate integral is
\be{ef1}
\II(U,T,V) = \int_{\FF} {d^2\tau\over \tau_2} \, Tr_{RR}\left(
(-1)^{(F_{K3} +
F_{T^2})}
(-1)^{(\bar F_{K3}+\bar F_{T^2})} F_{T^2} \bar F_{T^2}
q^{L_0} \bar q^{\bar L_0} 
\right)
\ee
where the trace is taken over the states in this orbifold superconformal 
field theory. 

For $V=0$ this integral has been calculated for the $\ZZZ_2$ orbifold
model in \cite{9708062}.  In the next few sections we shall describe
computation of this integral for the $N=2$ case for non-zero $V$.
This will enable us to determine the product form of $\wt\Phi_6$.
Later we shall discuss generalization of this analysis to other values
of $N$.

\section{The Integrand for the $\ZZZ_2$ Orbifold Theory} \label{s2}

In this section we shall analyze the integrand in eq.(\ref{ef1}) for
the $\ZZZ_2$ orbifold conformal field theory described earlier.  We
can decompose the contribution to the trace in (\ref{ef1}) as a sum of
the contribution from different sectors characterized by the five
charges $(m_1, n_1, m_2, n_2,b)$ introduced earlier.\footnote{Note
  that now the twisted sector states carry half integer winding number
  $n_1$ along $y^1$.}  In this case we can factor out the $T$, $U$ and
$V$ dependence of the trace into an overall factor of $q^{p_L^2/2}
\bar q^{p_R^2/2}$ where
\bea{e7}
{1\over 2} p_R^2 &=& {1\over 4 \det Im \Omega} |-m_1 U +
m_2 + n_1 T + n_2 (TU-V^2) + b V|^2, \nonumber \\
{1\over 2} p_L^2
&=& {1\over 2}  p_R^2 + m_1 n_1 + m_2 n_2 + {1\over 4} b^2\, .
\eea
Thus $\II(U,T,V)$ has the form
\be{ef2}
\II(U,T,V) = \int_{\FF} {d^2\tau\over \tau_2} \,
\sum_{m_1, m_2, n_1, n_2, b} q^{p_L^2/2-b^2/4} \bar
q^{p_R^2/2}
F_{m_1, m_2, n_1, n_2;b}(\tau)
\ee
where $F_{m_1, m_2, n_1, n_2;b}(\tau)$ is 
independent of $T$, $U$ and $V$ and 
is given by
\be{ef3}
F_{m_1, m_2, n_1, n_2;b}(\tau) = Tr_{m_1, m_2, n_1, n_2;b;RR}
\left( (-1)^{(F_{K3} +
F_{T^2})}
(-1)^{(\bar F_{K3}+\bar F_{T^2})} F_{T^2} \bar F_{T^2}
q^{L_0'} \bar q^{\bar L_0'}
\right)\, .
\ee 
Here
\be{ef4}
L_0'=L_0 - {p_L^2 \over 2}+{b^2\over 4}, 
\qquad \bar L_0'=\bar L_0 - {p_R^2\over 2} 
\, ,
\ee
are independent of $T$, $U$ and $V$ and $Tr_{m_1, m_2, n_1, n_2;b}$
denotes trace over a subspace of the Hilbert space carrying momentum
$(m_1,m_2)$ and winding $(n_1,n_2)$ along $T^2$ and $F_{K3}$ charge
$b$.  Note that we have included the $b^2/4$ term in $L_0'$ so that
for $V=0$ when the conformal field theories associated with $K3$ and
$T^2$ parts decouple, $L_0'$ and $\bar L_0'$ describe complete
contribution from the CFT associated with $K3$ and oscillator
contribution from the CFT associated with $T^2$.  Since $F_{m_1, m_2,
  n_1, n_2;b}(\tau)$ is independent of $T$, $U$ and $V$, we can set
$V=0$ while evaluating (\ref{ef3}).

Let us define
\be{e5} F_{m_1, m_2, n_1, n_2}(\tau,z) =\sum_{b}
F_{m_1, m_2, n_1, n_2;b}(\tau)
\, e^{2\pi i b z}\, .
\ee
It then follows from (\ref{ef3}) that
\be{ef6}
F_{m_1, m_2, n_1, n_2}(\tau,z) =  Tr_{m_1, m_2, n_1, n_2;RR}
\left( (-1)^{(F_{K3} +
F_{T^2})}
(-1)^{(\bar F_{K3}+\bar F_{T^2})} F_{T^2} \bar F_{T^2}
e^{2\pi i z F_{K3}} \, q^{L_0'} \bar q^{\bar L_0'}
\right)\, .
\ee
We shall first compute $F_{m_1, m_2, n_1, n_2}(\tau,z)$ and then extract 
$F_{m_1, m_2, n_1, n_2;b}(\tau)$ using eq.(\ref{e5}).
Since the contribution to (\ref{ef6}) from the $T^2$ part is
somewhat trivial, it is useful to separate out this contribution.
For this we denote by $g'$ the generator of the $\ZZZ_2$ group
with which we take the orbifold of $K3\times T^2$. Then
\bea{ef6new}
&& F_{m_1, m_2, n_1, n_2}(\tau,z) \nonumber \\
&=&  {1\over 2}\, \sum_{r,s=0}^1
\, Tr^{K3\times T^2}_{m_1, m_2, n_1, 
n_2;RR;(g')^r}
\left( (-1)^{(F_{K3} +
F_{T^2})}
(-1)^{(\bar F_{K3}+\bar F_{T^2})} F_{T^2} \bar F_{T^2}
e^{2\pi i z F_{K3}} \, q^{L_0'} \bar q^{\bar L_0'}
(g')^s \right)\, , \nonumber \\
\eea
where the superscript $K3\times T^2$ in $Tr$ indicates that the trace is 
taken in the superconformal field theory with target space $K3\times T^2$, 
and the subscript $(g')^r$ in $Tr$ indicates that the trace is over 
the sector twisted by $(g')^r$. We now
split $g'$ as
\be{esplit}
g' = \hat g \, \wt g\, ,
\ee
where $\hat g$ and $\wt g$ represent the action of $g'$ on the $T^2$
and $K3$ parts respectively. Twisting by $\hat g^r$ makes the winding
number $n_1\in \ZZZ+{r\over 2}$, and hence the right hand side of
(\ref{ef6new}) vanishes unless $n_1-{r\over 2}\in\ZZZ$. The $(\hat
g)^s$ factor inside the trace produces a factor of $(-1)^{m_1 s}$.
The non-zero mode bosonic and fermionic oscillator contributions from
the $T^2$ factor always cancel since they are neutral under $\hat g$.
The fermion zero modes associated with $T^2$ give a factor of 4 due to
2-fold degeneracy each from the holomorphic and anti-holomorphic
sectors, but this cancels with the factor of $1/4$ coming from the
$F_{T^2}\bar F_{T^2}$ factor inside the trace. Thus we can write
\be{ex3}
F_{m_1, m_2, n_1, n_2}(\tau, z) = \sum_{s=0}^1\, (-1)^{m_1\, s}\, 
F^{(r,s)}(\tau,z)  \qquad \hbox{for $n_1\in \ZZZ+{r\over 2}$}, 
\quad r=0,1
\ee
where
\be{esp1}
F^{(r,s)}(\tau,z) = {1\over 2} \, Tr^{K3}_{RR; \wt g^r}
\left( (-1)^{F_{K3}}
(-1)^{\bar F_{K3}} \, \wt g^s\, 
e^{2\pi i z F_{K3}} \, q^{L_0} \bar q^{\bar L_0}
\right)\, .
\ee
Here $Tr^{K3}_{RR; \wt g^r}$ denotes trace in the superconformal field
theory associated with target space $K3$ in the $\wt g^r$ twisted RR
sector, and $L_0$, $\bar L_0$ inside the trace now includes
contribution from $K3$ only.  This is twisted elliptic genus of $K3$.
These quantities were introduced in \cite{9306096} in order to
calculate the elliptic genus of $\wt g$ orbifold of $K3$. This would
be given by $\sum_{r,s=0}^{1} F^{(r,s)}(\tau,z)$. Here however we need
the individual $F^{(r,s)}(\tau, z)$ since we shall be using them for a
different purpose.

{}From the definitions given in (\ref{esp1}) it follows 
that\cite{9306096}
\be{ex2a}
F^{(r,s)}\left({a\tau+b\over c\tau + d}, {z\over c\tau + d}\right)
= \exp\left( 2\pi i {c\, z^2\over c\tau + d}\right) 
F^{(cs+ar, ds+br)}(\tau, z)\, ,
\ee
for
\be{ex2b}
a,b,c,d\in \ZZZ, \qquad ad - bc = 1\, .
\ee
In (\ref{ex2a}) the indices $cs+ar$ and $ds+br$ are to be taken mod 2.

$F_{m_1, m_2, n_1, n_2}(\tau, z)$
has been calculated in appendix \ref{sa} using an orbifold 
description of 
$K3$ and the result is
given in eq.(\ref{ee1}).  Comparing this with eq.(\ref{ex3})
we get
\bea{ex4}
&& F^{(0,0)}(\tau, z) = 4\left[ {\vartheta_2(\tau,z)^2
\over \vartheta_2(\tau,0)^2} +
{\vartheta_3(\tau,z)^2\over \vartheta_3(\tau,0)^2}
+ {\vartheta_4(\tau,z)^2\over \vartheta_4(\tau,0)^2}\right]\, ,
\nonumber \\ &&
F^{(0,1)}(\tau, z) =  4 \, {\vartheta_2(\tau,z)^2\over 
\vartheta_2(\tau,0)^2}\, , \quad
F^{(1,0)}(\tau, z) = 4 {\vartheta_4(\tau,z)^2\over \vartheta_4(\tau,0)^2}
\quad
F^{(1,1)}(\tau, z) =  4\, {\vartheta_3(\tau,z)^2\over 
\vartheta_3(\tau,0)^2}\, . \nonumber \\
\eea
Using the known modular transformation laws of 
$\vartheta_i(\tau, z)$ we can verify that $F^{(r,s)}(\tau,z)$ given in 
(\ref{ex4}) 
satisfy (\ref{ex2a}).

We now use the relations:
\bea{ee2}
\vartheta_1^2(\tau, z) &=& \vartheta_2(2\tau, 0) \vartheta_3(2\tau, 2z)
-\vartheta_3(2\tau, 0) \vartheta_2(2\tau, 2z) \nonumber \\
\vartheta_2^2(\tau, z) &=& \vartheta_2(2\tau, 0) \vartheta_3(2\tau, 2z)
+\vartheta_3(2\tau, 0) \vartheta_2(2\tau, 2z) \nonumber \\
\vartheta_3^2(\tau, z) &=& \vartheta_3(2\tau, 0) \vartheta_3(2\tau, 2z)
+\vartheta_2(2\tau, 0) \vartheta_2(2\tau, 2z) \nonumber \\
\vartheta_4^2(\tau, z) &=& \vartheta_3(2\tau, 0) \vartheta_3(2\tau, 2z)
-\vartheta_2(2\tau, 0) \vartheta_2(2\tau, 2z)  
\eea
to rewrite (\ref{ex4}) as
\be{ee3}
F^{(r,s)}(\tau, z) = h_0^{(r,s)}(\tau) 
 \, \vartheta_3(2\tau, 2z)
+ h^{(r,s)}_1(\tau) 
\, \vartheta_2(2\tau, 2z) 
\ee
where
\bea{ee4}
&& h^{(0,0)}_0(\tau) = 8\, {\vartheta_3(2\tau,0)^3 \over \vartheta_3(\tau,0)^2
\vartheta_4(\tau,0)^2} + 2\, {1\over \vartheta_3(2\tau,0)}
\nonumber \\
&& h^{(0,0)}_1(\tau) = -8\, {\vartheta_2(2\tau,0)^3 
\over  \vartheta_3(\tau,0)^2
\vartheta_4(\tau,0)^2} + 2\, {1\over \vartheta_2(2\tau,0)}
\nonumber \\
&& h_0^{(0,1)}(\tau) = 2\, {1\over \vartheta_3(2\tau,0)}, \qquad
h_1^{(0,1)}(\tau) = 2\, {1\over \vartheta_2(2\tau,0)},
\nonumber \\
&& h_0^{(1,0)}(\tau) = 4 {\vartheta_3(2\tau, 0)\over
\vartheta_4(\tau,0)^2}, \qquad
h_1^{(1,0)}(\tau) =  -4 {\vartheta_2(2\tau, 0)\over
\vartheta_4(\tau,0)^2},
\nonumber \\
&& h_0^{(1,1)}(\tau) = 4 {\vartheta_3(2\tau, 0)\over
\vartheta_3(\tau,0)^2}, \qquad
h_1^{(1,1)}(\tau) =  4 {\vartheta_2(2\tau, 0)\over
\vartheta_3(\tau,0)^2},
\eea
Since
\be{eg1}
\vartheta_3(2\tau, 2 z) = \sum_{b\in 2\zzz} e^{2\pi i b z} q^{b^2/4}, 
\qquad \vartheta_2(2\tau, 2 z) = \sum_{b\in 2\zzz+{1}} e^{2\pi i b z} 
q^{b^2/4},
\ee
we see, by comparing (\ref{e5}) and (\ref{ex3}), (\ref{ee3}) 
that 
\bea{eg2}
F_{m_1, m_2, n_1, n_2;b}(\tau) &=& q^{b^2/4}\, 
\sum_{s=0}^1\,
(-1)^{m_1 s} \, h^{(r,s)}_l(\tau)
\quad 
\hbox{for $n_1\in \ZZZ+{r\over 2}$, $b\in 
2\ZZZ+l$} \nonumber \\
&& \qquad \qquad \qquad \qquad \quad r,l=0, 1\, .
\eea
Using (\ref{eg2}) the original integral $\II(U,T,V)$ 
given in eq.(\ref{ef2}) may be written as
\be{eh1}
\II(U,T,V) = \sum_{l,r,s=0}^1
\, \II_{r,s,l} 
\ee
where
\be{eh2}
\II_{r,s,l} = \int_{\FF} {d^2\tau\over \tau_2} \,
\sum_{m_1, m_2, n_2 \in \zzz\atop
n_1\in \zzz+{r\over 2},
b\in 2\zzz +l} q^{p_L^2/2} \bar
q^{p_R^2/2} \, (-1)^{m_1 s}\, 
h^{(r,s)}_l(\tau)\, .
\ee
{}From this we see that those $SO(2,3;\ZZZ)$ transformations which,
acting on a vector $(m_1,m_2,n_1, n_2,b)$ with $m_1,m_2,n_2, b$
integers and $n_1$ half-integer, preserves $m_1$ modulo 2,
$n_1,m_2,n_2$ modulo 1 and $b$ modulo 2, will be symmetries of $\III$.
This defines the subgroup $\wt G$.

For later use we define the coefficients $c^{(r,s)}(4n)$ 
through the expansion
\be{ee5}
h^{(r,s)}_0(\tau) = \sum_{n  } 
c^{(r,s)}(4n) 
q^n\, , \qquad h^{(r,s)}_1(\tau)
= \sum_{n  } c^{(r,s)}(4n) q^n\, .
\ee
By examining (\ref{ee4}) we see that in the expansion of
$h^{(r,s)}_l$, $n\in \ZZZ-{l\over 4}$ for $r=0$ and $n\in {1\over 2}
\ZZZ-{l\over 4}$ for $r=1$. Note that we have used the same symbol
$c^{(r,s)}(4n)$ for describing the expansion of $h^{(r,s)}_0(\tau)$
and $h^{(r,s)}_1(\tau)$.  This is possible since $c^{(r,s)}(4n)$ has
different support for $l=0$ and $l=1$.

Using eq.(\ref{ee3}) and the Fourier expansion (\ref{eg1}) of
$\vartheta_3$ and $\vartheta_2$ we can write the double Fourier
expansion of $F^{(r,s)}(\tau, z)$ 
\be{efourier} 
F^{(r,s)}(\tau,z)=\sum_{b\in \zzz, n } c^{(r,s)}(4n-b^2)\, 
q^{n} \, e^{2\pi i z b}\, ,
\ee 
where $n\in \ZZZ$ for $r=0$ and ${1\over 2}\ZZZ$ for $r=1$.

\section{The Integral} \label{s3}

We shall now proceed to evaluate the integral (\ref{eh2}).
We define
\be{defmm}
Y = \det {\rm Im} \Omega = T_2 \, U_2 - 
(V_2)^2, \qquad
 T_2>0 ,\quad   U_2>0, \quad 
Y > 0\, .
\ee
where for any complex number $u$, we denote
by $u_1$ and $u_2$ its real and imaginary parts respectively.
Substituting the values of $p_L^2$ and $p_R^2$ 
from \eq{e7} into
\eq{eh2} we obtain
\bea{bint1}
\II_{r,s,l} &=& \int_\FF
\frac{d^2\tau}{\tau_2} \sum_{m_1, m_2, n_2\in \zzz, 
n_1\in \zzz+{r\over 2}, b\in 2\zzz + l}
\exp\left[ 2\pi i \tau( m_1 n_1 + m_2 n_2 +\frac{b^2}{4} ) \right]
\times \cr
&\;& \;\;\; \exp \left[\frac{-\pi \tau_2}{Y} \left|
n_2 ( TU -V^2) + bV + n_1 T -Um_1 + m_2 \right|^2 \right]\, 
(-1)^{m_1 s} \, h^{(r,s)}_l(\tau)\, . \nonumber \\
\eea
To evaluate the integral we first perform the Poisson resummation 
over the momenta $m_1, m_2$.
The basic formula for Poisson resummation we will use is
\be{prsum}
\sum_{m\in \zzz} f(m) e^{2\pi i s m/N} 
= \sum_{k\in \zzz+{s\over N}} 
\int_{-\infty}^{\infty} du f(u) \exp( 2\pi i k u)
\ee
for any integer $N$.
Now performing the Poisson resummation over $m_1, m_2$ and
performing the Gaussian integration over the corresponding
variables $u_1$, $u_2$, we obtain 
the following 
\be{bint3}
\II_{r,s,l} = \int_\FF \frac{d^2\tau}{\tau_2^2} \frac{Y}{U_2} 
\, \sum_{n_2, k_2\in \zzz, n_1\in \zzz+{r\over 2}, k_1\in \zzz+{s\over 
2}, b\in 2\zzz+l} \, h^{(r,s)}_l(\tau)
\, \exp\left[ {\cal G}(\vec n, \vec k, b)\right]
\ee
where
\bea{fing}
{\cal G}(\vec n, \vec k, b) &=& - 
\frac{\pi Y}{U_2^2 \tau_2}|{\cal A}|^2 
- 2\pi i T {\rm det}\,  A \cr
&+& \frac{\pi b}{U_2} ( V\tilde {\cal A} - \bar V {\cal A})
- \frac{\pi n_2}{U_2} ( V^2 \tilde{\cal A} - \bar V^2 {\cal A})
\cr
&+& \frac{2 \pi i V_2^2}{U_2^2} ( n_1 + n_2 \bar U) {\cal A}
 + 2\pi i \tau \frac{b^2}{4}\, ,
\eea
\be{fing1}
A = \left(
\begin{array}{cc}
n_1 & k_1 \\
n_2 & k_2 
\end{array}
\right)\, ,
\ee
\be{fing2}
{\cal A} = ( 1 , U) A 
\left(
\begin{array}{c}
\tau \\ 1
\end{array}
\right)\, ,
\qquad
\tilde {\cal A} = ( 1 , \bar U) A 
\left(
\begin{array}{c}
\tau \\ 1
\end{array}
\right)\, .
\ee
Using (\ref{fing}) we can represent the sum over $b$ in (\ref{bint3}) as
\be{ey1}
\sum_{b\in 2\zzz+l} e^{2\pi i \tau \frac{b^2}{4} + \frac{\pi b}{U_2} ( 
V\tilde {\cal A} - \bar V {\cal A}) }
= \cases{\vartheta_3(2\tau, -i { V\tilde {\cal A} - \bar V {\cal A} \over 
U_2}) \quad \hbox{for $l=0$} \cr
\vartheta_2(2\tau, -i { V\tilde {\cal A} - \bar V {\cal A} \over
U_2}) \quad \hbox{for $l=1$} }
\ee
Substituting this into (\ref{bint3}) and using (\ref{ee3})
we get
\be{ey2}
\II \equiv \sum_{l,r,s=0}^1 = 
\int_\FF
\frac{d^2\tau}{\tau_2^2} \, \sum_{r,s=0}^1\, 
\sum_{n_2, k_2\in \zzz, n_1\in \zzz+{r\over 2}, 
k_1\in \zzz+{s\over
2}} \JJ(A,\tau)\, ,
\ee
where
\bea{ezz2}
\JJ(A, \tau) &=& \frac{Y}{U_2}
\, 
\exp\Bigg(  - \frac{\pi Y}{U_2^2 \tau_2}|{\cal A}|^2
- 2\pi i T {\rm det}\,  A \nonumber \\
&& - \frac{\pi n_2}{U_2} ( V^2
\tilde{\cal A} -
\bar V^2 {\cal A}) + \frac{2 \pi i V_2^2}{U_2^2} ( n_1 + n_2 \bar U) {\cal
A} \Bigg) \, F^{(r,s)}\left(\tau, -i { V\tilde {\cal A} -
\bar V
{\cal A} \over
2\, U_2}\right) \nonumber \\
&&  \qquad \qquad r =2\, n_1 \, \hbox{mod 2}, \quad s = 2 
\, k_1 \, \hbox{mod 2}\, .
\eea
In order to interpret the right hand side as a function of the matrix $A$
we need to use eqs.(\ref{fing1}), (\ref{fing2}). We may now interpret
the sum over $r,s$ and $\vec n, \vec k$ in the right hand side of
eq.(\ref{ey2}) as a sum over all matrices $A$ of the form (\ref{fing1})
with $n_2$, $k_2$ integer, and $n_1$, $k_1$ integer or half-integer.
(\ref{ey2}) may then be rewritten as
\be{ezz3}
\II =  \int_\FF
\frac{d^2\tau}{\tau_2^2} \,
\sum_A \, \JJ(A, \tau)\, .
\ee

Now it follows from the modular transformation laws (\ref{ex2a}) and the 
definition of $\JJ(A, \tau)$ given in (\ref{ezz2}) that
\be{ezz4}
\JJ\left(A, {a\tau+b\over c\tau + d}\right) = 
\JJ \left( A \pmatrix{a & b\cr c & d}, \tau\right)\, .
\ee
Using this symmetry, we can extend the integration over the
fundamental domain to its images under $SL(2,\ZZZ)$ and at the same
time restrict the summation over $A$ to summation over inequivalent
$SL(2,\ZZZ)$ orbits.  If we denote by $\sum'_A$ the sum over
inequivalent $SL(2,\ZZZ)$ orbits then we can express $\II$ as
\bea{ezz5}
\II &=& \sum'_A 
 \int_{\FF_A}
\frac{d^2\tau}{\tau_2^2} \,
\frac{Y}{U_2}
\, \exp\Bigg(  - \frac{\pi Y}{U_2^2 \tau_2}|{\cal A}|^2
- 2\pi i T {\rm det}\,  A \nonumber \\
&& - \frac{\pi n_2}{U_2} ( V^2
\tilde{\cal A} -
\bar V^2 {\cal A}) + \frac{2 \pi i V_2^2}{U_2^2} ( n_1 + n_2 \bar U) {\cal
A} \Bigg) \, F^{(r,s)}\left(\tau, -i { V\tilde {\cal A} -
\bar V
{\cal A} \over
2\, U_2}\right) \nonumber \\
\eea
where now $r$, $s$ in the label of $F^{(r,s)}$ are to be interpreted
as $2\, n_1$ mod 2 and $2 \, k_1$ mod 2 respectively. The region of
integration $\FF_A$ depends on the orbit represented by $A$.

Following the same procedure as in \cite{dixon} we now split the
integration into the three orbits. These are
the zero orbit
\be{ezero} A = 0\, ,
\ee
the non-degenerate orbit
\be{nondeag}
A = \left(
\begin{array}{cc}
k & j \\
0 & p
\end{array}
\right), \quad 2k-1\geq 2j \geq 0, \quad p\neq 0, \quad k,j\in {1\over 2}
\ZZZ, \quad
p\in  \ZZZ \, ,
\ee
and
the degenerate orbit
\be{degaag}
A = \left(
\begin{array}{cc}
0 & j \\
0 & p
\end{array}
\right), \quad (j, p) \neq (0,0), \quad j\in {1\over 2}
\ZZZ, \quad
p\in  \ZZZ\, .
\ee
The contribution from these orbits has been evaluated 
in appendix \ref{sb}. The final result, as given in (\ref{finalfulk3app}), 
takes the form
\bea{finalfulk3}
 {\cal I}  &=& - 2 \ln \Bigg[
{\kappa} (\det{\rm Im}\Omega)^{6} \Bigg|
\exp\left( 2\pi i \left( {1\over 2}T+ U+ V\right) 
\right) \nonumber \\
&& \prod_{r,s=0}^1\, \prod_{(l,b)\in \zzz, k\in \zzz+{r\over 2}
\atop
(k, l, b)>0} \, \Bigg\{
( 1 - \exp( 2\pi i ( k T + l U + b V) ) ^{(-1)^{ls}\, 
c^{(r,s)}(4kl - b^2)}  \Bigg\} \Bigg |^2 \Bigg] 
 \cr
\kappa &=& \left ( \frac{8\pi}{ 3\sqrt 3 } e^{1-\gamma_E} \right)^{6}
\eea
and $(k, l , b)>0$ means $k> 0 , l\geq 0, b \in \ZZZ$ or 
$ k =0, l>0, b \in \ZZZ$
or $k =0, l =0, b <0$. 

\section{$\wt\Phi_6$ and its $V\to 0$ Limit} \label{s3.5}

Eq.(\ref{finalfulk3}) can be written as
\be{erewrite}
\II = -2\left[ 6\ln\det{\rm Im}\Omega + \ln\wt\Phi_6+ \ln \bar
{\wt\Phi}_6 + 
\ln\kappa + 8\ln 2\right]\, ,
\ee
where
\bea{defpi10}
\wt\Phi_{6}(\Omega) &=& {1\over 16}
\, \exp\left(2\pi i \left( {1\over 2}\,
T+ U + V\right) \right) \nonumber \\
&&
\prod_{r,s=0}^1\, \prod_{l,b\in \zzz, k\in \zzz+{r\over 2}\atop
 k,l,b>0}  \bigg[
 1 - \exp\left\{ 2\pi i( kT + l U + bV)\right\} \bigg]^{
(-1)^{ls} c^{(r,s)}(4lk -b^2)
  }    \, .
\eea
Note that we have normalized $\wt\Phi_6$ so that the coefficient of
$\exp(2\pi i ( {1\over 2}\, T+ U + V) )$ is 1/16. This agrees with the
normalization convention of \cite{0510147}.

Since under a duality transformation by an element
$\pmatrix{A & B\cr C & D}$ of
$\wt G\subset Sp(2,\ZZZ)$  
\be{transy}
\det{\rm Im}\Omega \rightarrow |{\rm det}(C\Omega + D)|^{-2} 
\det{\rm Im}\Omega\, ,
\ee
we must have
\be{ephiweight}
\wt\Phi_6((A\Omega+B)(C\Omega+D)^{-1})
= \det(C\Omega+D)^6 \wt\Phi_6(\Omega)\, ,
\ee
in order that $\II$ given in (\ref{erewrite}) is invariant under this
transformation.
Thus $\wt\Phi_6$ transforms as a modular form of weight 6
under $\wt G$. 

We shall now analyze the $V\to 0$ limit of (\ref{defpi10}) and compare
this with the corresponding result in \cite{0510147}.  This analysis
is facilitated by examining the relation (\ref{efourier}) at $z=0$:
\be{ecexpan}
\sum_n \sum_b c^{(r,s)}(4n - b^2) q^n = F^{(r,s)}(\tau, 0)
= \cases{12 \quad \hbox{for}\quad 
(r,s)=(0,0)\cr 4 \, \, \, \quad \hbox{for}\quad
(r,s)\ne (0,0)}\, .
\ee
This gives
\be{ecx1}
\sum_b c^{(r,s)}(4n - b^2) = \cases{12 \, \delta_{n,0}
\quad \hbox{for}\quad 
(r,s)=(0,0)\cr 4 \, \delta_{n,0} \, \, \, \quad \hbox{for}\quad
(r,s)\ne (0,0)}\, .
\ee
Taking $V\to 0$ limit in (\ref{defpi10}) we now get
\bea{ecx2pre}
\wt\Phi_6(U,T,V) &\simeq& -{4\pi^2 V^2\over 16} \,
e^{2\pi i\left({1\over 2}T+U\right)} \prod_{k=1\atop 
k\in\zzz}^\infty 
\left\{\left( 1 - e^{2\pi i k T}\right)^{8}
\left( 1 - e^{\pi i k T}\right)^{8}\right\} \nonumber \\
&& \prod_{l=1\atop l\in\zzz}^\infty 
\left\{\left( 1 - e^{2\pi i l U}\right)^{8}
\left( 1 - e^{4\pi i l U}\right)^{8}\right\}
\eea
where the $-4\pi^2 V^2$ term comes from the $k=l=0$, $b=-1$
term. This can be rewritten as 
\be{ecx2}
 \wt\Phi_6(U,T,V) \simeq
 -{1\over 4}\, \pi^2\, V^2  \eta(T/2)^8 \eta(T)^8 \eta(U)^8 \eta(2U)^8
 \, .
 \ee
This factorization property, including the overall 
normalization of $-{1\over 4}\pi^2$, agrees with that found in 
\cite{0510147}.

\section{Construction of $\Phi_6$} \label{s4}

In the analysis of \cite{0510147} we introduced another function
$\Phi_6$ related to $\wt\Phi_6$ by:
\be{ei2.4}
   \wt\Phi_6(U,T, V)
   = T^{-6}
   \Phi_6\left(
U -{V^2\over T}, -{1\over T},
     {V\over T}\right)\, ,
\ee
or equivalently
\be{ei2.5}
   \Phi_6(U,T, V)
   = T^{-6}
   \wt\Phi_6\left(
U -{V^2\over T},-{1\over T},
     {V\over T}\right)\, .
\ee
{}From the expressions for $\II_{r,s,l}$ given in (\ref{bint1})
we see that this transformation may be implemented by
\be{ei2.6}
m_2\to n_1, \quad n_1\to -m_2, \quad m_1\to -n_2, 
\quad n_2\to m_1\, .
\ee
Thus in order to find an expression for $\Phi_6$ we can replace
$\II_{r,s,l}$ given in (\ref{bint1}) by $\II'_{r,s,l}$ in which we sum
over $m_2\in \ZZZ+{r\over 2}$ instead of $n_1\in \ZZZ+{r\over 2}$, and
replace the $(-1)^{m_1 s}$ factor in the summand by $(-1)^{n_2 s}$:
\bea{bint1til}
\II'_{r,s,l} &=& \int_\FF
\frac{d^2\tau}{\tau_2} \sum_{m_1, n_1, n_2\in \zzz,
m_2\in \zzz+{r\over 2}, b\in 2\zzz + l}
\exp\left[ 2\pi i \tau( m_1 n_1 + m_2 n_2 +\frac{b^2}{4} ) \right]
\times \cr
&\;& \;\;\; \exp\left[\frac{-\pi \tau_2}{Y} \left|
n_2 ( TU -V^2) + bV + n_1 T -Um_1 + m_2 \right|^2 \right]\,
(-1)^{n_2 s} \, h^{(r,s)}_l(\tau)\, . \nonumber \\
\eea
After Poisson resummation this amounts to 
summing over only integer values of $n_1$, $n_2$, $k_1$, $k_2$ and 
including a factor of 
\be{ezy1}
(-1)^{k_2 r} (-1)^{n_2 s} \, ,
\ee
in the summand. The integral can now be evaluated following exactly
the same procedure as in appendix \ref{sb}, the only difference being
that the sum over $p$ in eqs.(\ref{finint2}), (\ref{suminfo}),
(\ref{sbint3}) will contain an additional factor of $(-1)^{p\,
  r}$.\footnote{An apparent additional complication arises due to the
  fact that the Fourier expansions of $F^{(1,0)}$ and $F^{(1,1)}$ as
  given in (\ref{efourier}) have half integer powers of $q$. Thus the
  sum over $j$ in eq.(\ref{jdep}) will not vanish for non-integer
  $n/k$. However since $F^{(1,0)}+F^{(1,1)}$ is invariant under
  $\tau\to\tau+1$ due to the modular properties described in
  (\ref{ex2a}), it has Fourier expansion in integer powers of $q$.
  Thus if in analyzing the sum over $j$ in (\ref{jdep}) we consider
  the contribution from $F^{(1,0)}$ and $F^{(1,1)}$ together, the sum
  over $j$ will force $n$ to be a multiple of $k$.}  The net
contribution to the full integral comes out to be
\bea{enetcont}
{\cal I'} &=& - 2 \ln \Bigg[2^8\, 
\kappa (\det{\rm Im}\Omega)^{6} \Bigg|
\exp( 2\pi i (T+ U+ V) ) \nonumber \\
&& \prod_{r,s=0}^1\, \prod_{(k,l.b)\in \zzz\atop
(k, l, b)>0}
\Big\{ 1 - (-1)^r \exp( 2\pi i ( k T + l U + b V) 
\Big\}^{c^{(r,s)}(4kl - b^2)} 
\Bigg |^2 \Bigg]\, .
\nonumber \\
\eea
We can rewrite this as 
\be{erewritenew}
\II' = -2\left[ 6\ln\det{\rm Im}\Omega + \ln\Phi_6+ \ln \bar
{\Phi}_6 +
\ln\kappa+8\ln 2\right]\, ,
\ee
where
\bea{defpi10new}
\Phi_{6}(\Omega) &=& -\exp(2\pi i (
T+ U + V) ) \nonumber \\
&& \prod_{r,s=0}^1\, \prod_{(k,l.b)\in \zzz\atop 
(k, l, b)>0}
\Big\{ 1 - (-1)^r\, 
\exp( 2\pi i ( k T + l U + b V) \Big\}^{ c^{(r,s)}(4kl - b^2)}  \, .
\nonumber \\
\eea
The normalization of $\Phi_6$ is not arbitrary; it has been chosen so
that we have the same additive constant $8\ln 2$ in
(\ref{erewritenew}) as in (\ref{erewrite}). The phase of $\Phi_6$ can
be adjusted.  With the choice of phase given in (\ref{defpi10new}) the
coefficient of the $e^{2\pi i(T+U+V)}$ term matches with that of the
corresponding expression in \cite{0510147}.  Following the same
argument as in the case of $\wt\Phi_6$ we can argue that $\Phi_6$
transforms as a modular form of weight 6 under a subgroup $G$ of
$Sp(2,\ZZZ)$ which is related to the earlier subgroup $\wt G$ by the
conjugation described in (\ref{ei2.6}).

Study of the $V\to 0$ limit of this expression is also straightforward. 
Using the relations (\ref{ecx1}) and the explicit expressions for the 
coefficients $c^{(r,s)}(0)$ and $c^{(r,s)}(-1)$ given in (\ref{ecvalue}), 
we get
\be{ephi6v}
\Phi_6(U,T, V) \simeq 4\pi^2\, V^2\,
\eta(T)^8 \eta(2T)^8 \eta(U)^8 
\eta(2U)^8\, .
\ee
This is the same behaviour as found in \cite{0510147}.

We can also carry out a more detailed comparison between
the $\Phi_6$ defined here and those in \cite{0510147}.
The algorithm given in \cite{0510147} goes as follows:
\begin{itemize}
\item We first define a set of coefficients $f_n$ ($n\ge 1$) through
the relation:
 \be{ei1}
 \sum_{n\ge 1} f_n e^{2\pi i \tau
     (n-{1\over 4})} =  \eta(\tau)^{2}\, \eta(2\tau)^{8}\, ,
\ee
where $\eta(\tau)$ is the Dedekind  function.  

\item Next we define the coefficients $C(m)$ through 
\bea{eisk2} C(m) = (-1)^m 
\sum_{s,n\in\zzz \atop n\ge 1} f_n
\delta_{4n+{s^2-1 },{m }}\, . 
\eea 

\item $\Phi_6$ is now given by
\be{eisk3}
   \Phi_6(U,T,V) = \sum_{n,m,r\in\zzz\atop
     n,m\ge 1, \, r^2< 4mn}\, a(n,m,r)
   \, e^{2\pi i (nU+mT+rV)}
   \, ,
\ee
where
\be{eisk4}
   a(n, m,r) = \sum_{\alpha\in 2\zzz+1\atop
     \alpha|(n,m,r), \alpha>0} \,  
   \alpha^{k-1}\, C\left({4mn
       -r^2\over \alpha^2}\right)\, ,
\ee

\end{itemize}

We have compared 31 different coefficients $a(n,m,r)$ defined in
(\ref{eisk4}) with the ones obtained from (\ref{defpi10new})
and found them to be the same. These results for $a(n,m,r)$
are given in appendix \ref{sc}.

\section{Construction of $\Phi_k$ and $\wt\Phi_k$} \label{s5}

Generalization of the modular form $\wt\Phi_6$ to describe the
degeneracy of dyons in a $\ZZZ_N$ orbifold of $T^2\times K3$ for
$N=2,3,5,7$ was also introduced in \cite{0510147}.  The generator $g'$
of the $\ZZZ_N$ is given by
\be{egenzn}
g' = \hat g\, \wt g\, ,
\ee
where $\hat g$ represents $1/N$ unit of shift along $T^2$ (which we
shall take to be in the $y^1$ direction) and $\wt g$ denotes an
appropriate $\ZZZ_N$ action on $K3$. $\wt g$ preserves the harmonic
(0,0)-form, (2,2)-form, (0,2)-form and (2,0)-form. Furthermore for
each $r\ne 0$, there are $24/(N+1)$ (1,1)-forms with $\wt g$
eigenvalue $e^{2\pi i r/N}$. The rest of the $20 - 24 (N-1)/ ( N+1)$
of the (1,1)-forms are invariant under $\wt g$.

The generating function for the 
degeneracy is given by $(\wt\Phi_k)^{-1}$ where 
\be{edefk}
k = {24\over N+1}-2\, ,
\ee
and $\wt\Phi_k$ is a weight $k$ modular form of a subgroup $\wt G$ of 
$Sp(2,\ZZZ)=SO(2,3;\ZZZ)$ that commutes with $1/N$ unit of shift along a 
circle of $T^2$. Associated with $\wt\Phi_k$ there is a modular form 
$\Phi_k$ of a different subgroup $G$ of $Sp(2,\ZZZ)$, related to $G$ by 
conjugation described in  (\ref{ei2.5}):
\be{ei2.5new}
   \Phi_k(U,T, V)
   = T^{-k}
   \wt\Phi_k\left(
U -{V^2\over T},-{1\over T},
     {V\over T}\right)\, .
\ee

Our goal is to find a product representation of $\Phi_k$ and
$\wt\Phi_k$.  For this we shall start with an analog of eq.(\ref{ef1})
for the superconformal field theory associated with the $\ZZZ_N$
orbifold of $K3\times T^2$ and express it as a sum of a holomorphic
and an anti-holomorphic term and a term proportional to $\ln\det{\rm
  Im}\Omega$.  The holomorphic part can then be identified with
$\Phi_k$.  Proceeding as in section \ref{s1} we arrive at the analog
of eq.(\ref{ex3}), (\ref{esp1})
\be{ex3new}
F_{m_1, m_2, n_1, n_2}(\tau, z) = \sum_{s=0}^{N-1}
\, e^{2\pi im_1 s/N} \,\,
F^{(r,s)}(\tau,z)  \qquad \hbox{for $n_1\in \ZZZ+{r\over N}$},
\quad r=0,1, \ldots (N-1)\, ,
\ee
where
\be{esp1new}
F^{(r,s)}(\tau,z) = {1\over N} \, Tr^{K3}_{RR; \wt g^r}
\left( (-1)^{F_{K3}}
(-1)^{\bar F_{K3}} \, \wt g^s\,
e^{2\pi i z F_{K3}} \, q^{L_0} \bar q^{\bar L_0}
\right)\, .
\ee
{}From these definitions
it follows that
\be{ex2anew}
F^{(r,s)}\left({a\tau+b\over c\tau + d}, {z\over c\tau + d}\right)
= \exp\left( 2\pi i {c z^2\over c\tau + d}\right)
F^{(cs+ar, ds+br)}(\tau, z)\, ,
\ee
for
\be{ex2bnew}
a,b,c,d\in \ZZZ, \qquad ad - bc = 1\, .
\ee
In (\ref{ex2anew}) the indices $cs+ar$ and $ds+br$ are to be 
taken mod N. Thus for each $(r,s)$, $F^{(r,s)}(\tau, z)$ transforms
as a weak Jacobi form\cite{eichler} 
of weight zero and index 1 under the group
$\Gamma(N)$.

We can now define the coefficients $c^{(r,s)}(n)$ in a manner
analogous to (\ref{efourier})\footnote{In order that
  $F^{(r,s)}(\tau,z)$ has an expansion of the form given in
  (\ref{efouriernew}) we need to ensure that this can be expressed as
  a linear combination of $\vartheta_3(2\tau, 2z)$ and
  $\vartheta_2(2\tau, 2z)$ with $z$-independent coefficients as in
  (\ref{ee3}). This follows from the fact that the $z$-dependence of
  $F^{(r,s)}(\tau,z)$ comes from the SU(2) current algebra associated
  with the superconformal field theory, and this commutes with the
  $\ZZZ_N$ generator $\wt g$.  $\vartheta_3(2\tau, 2z)$ and
  $\vartheta_2(2\tau,2z)$ simply represent the contributions from the
  even and odd $F_{K3}$ charge sector of this SU(2) sector of the
  theory.  }
\be{efouriernew}
F^{(r,s)}(\tau,z) =\sum_{b\in \zzz, n\in\zzz/N }
c^{(r,s)}(4n-b^2)\,  q^{n} \,
e^{2\pi i z b}\, .
\ee
Contribution to $c^{(0,s)}(l)$ for $l=0$, $-1$ comes from geometric
data of $K3$ and can be computed easily. In particular untwisted
sector states with $n=0$, $b=0$ are associated with (1,1)-forms, those
with $n=0$, $b=1$ are associated with the (2,2) and the (2,0)-forms,
and those with $n=0$, $b=-1$ are associated with the (0,0) and the
(0,2)-forms.  Thus $N c^{(0,s)}(0)$ measures the trace of $\wt g^s$ on
the (1,1)-forms of $K3$ and $N c^{(0,s)}(-1)$ measures the trace of
$\wt g^s$ on the (0,0), (0,2) or (2,0), (2,2)-forms of $K3$.  These
can be easily computed from the $\wt g$ action of the cycles described
earlier, and we get
\bea{ecrsv}
&& c^{(0,0)}(0) = {20\over N}\, , \qquad 
c^{(0,0)}(-1) = {2\over N}\, , \nonumber \\
&& c^{(0,s)}(0) = {1\over 
N}\left( 20 - {24 
N\over N+1}\right)\, , 
\qquad c^{(0,s)}(-1) = {2\over N}\, , \quad \hbox{for 
$s=1,2, \ldots (N-1)$}\, . \nonumber \\
\eea
Several other useful properties of $c^{(r,s)}$ may be derived without
explicitly computing $F^{(r,s)}(\tau,z)$. 
First note that $F^{(0,0}(\tau, z)$ is $1/N$ times the elliptic genus
of $K3$. Hence it is given by
\be{esz0}
F^{(0,0)}(\tau, z) = {8\over N}\left[ {\vartheta_2(\tau,z)^2
\over \vartheta_2(\tau,0)^2} +
{\vartheta_3(\tau,z)^2\over \vartheta_3(\tau,0)^2}
+ {\vartheta_4(\tau,z)^2\over \vartheta_4(\tau,0)^2}\right]\, .
\ee
Next  it follows from the
definition (\ref{esp1new}) that $F^{(0,s)}(\tau, 0)$ is $\tau$
independent since it receives contribution only from the 
$L_0=\bar L_0=0$ states. The modular transformation laws
(\ref{ex2anew}) together with (\ref{ecrsv}) then imply that
\bea{esz1}
F^{(r,s)}(\tau, 0) = F^{(0, t)}(\tau, 0)|_{t=g.c.d.(r,s)} 
&=& c^{(0,t)}(0)
+ 2 \, c^{(0,t)}(-1) = {24 \over N(N+1)} \nonumber \\
&&
\hbox{for} \quad (r,s)\ne (0,0)\, .
\eea
Substituting (\ref{esz0}), (\ref{esz1}) into the expansion
(\ref{efouriernew}) we get the analog of eq.(\ref{ecx1})
\be{esz2}
\sum_b c^{(r,s)}(4n - b^2) = \cases{{24\over N} \, \delta_{n,0}
\quad \hbox{for}\quad 
(r,s)=(0,0)\cr {24 \over N(N+1)}\, \delta_{n,0} 
\, \, \, \quad \hbox{for}\quad
(r,s)\ne (0,0)}\, .
\ee
Further information about these coefficients comes from the fact that
$\sum_{r,s=0}^{N-1}F^{(r,s)}(\tau, z)$ represent the elliptic genus of
the super-conformal $\sigma$-model with target space $K3/\ZZZ_N$ with
the $\ZZZ_N$ generated by $\wt g$. However for any $N$ this gives us
back the superconformal field theory with target space $K3$, and hence
$\sum_{r,s=0}^{N-1}F^{(r,s)}(\tau, z)$ must give us the elliptic genus
of $K3$. This in turn is just $NF^{(0,0)}(\tau,z)$.  Thus we have
\be{esz4}
\sum_{r,s=0}^{N-1}F^{(r,s)}(\tau, z) = N \, F^{(0,0)}(\tau,z)\, .
\ee
Furthermore the contribution $\sum_{s=0}^{N-1}F^{(r,s)}(\tau, z)$ for
a fixed $r$ may be interpreted as the contribution to the elliptic
genus from the sector twisted by $\wt g^r$. For prime values of $N$,
$\wt g^r$ is an order $N$ transformation for all $r\ne 0$ mod $N$.
Hence we expect the sectors twisted by $\wt g^r$ to give the same
contribution to the elliptic genus for all $r\ne 0$ mod $N$. This,
together with (\ref{esz4}), gives
\be{esz5}
\sum_{s=0}^{N-1}F^{(r,s)}(\tau, z) = {1\over N-1}\left[
N \, F^{(0,0)}(\tau,z) - 
\sum_{s=0}^{N-1}F^{(0,s)}(\tau, z) 
\right]\,  \qquad r\ne \hbox{0 mod $N$}\, .
\ee
Translated to a condition on the
coefficients $c^{(r,s)}(m)$, this gives
\be{esz3}
\sum_{s=0}^{N-1}c^{(r,s)}(m) = {1\over N-1}\left[
N \, c^{(0,0)}(m) - 
\sum_{s=0}^{N-1}c^{(0,s)}(m) 
\right]\,  \quad \hbox{ for any $m$, \quad $r\ne 0$ mod $N$}\, .
\ee
For $m=0,-1$ we can explicitly evaluate the right hand side of this
equation using (\ref{ecrsv}). In particular setting $m=-1$ we get
\be{esz6}
\sum_{s=0}^{N-1}c^{(r,s)}(-1) = 0\, , \qquad \hbox{for $r\ne 0$
mod $N$}\, .
\ee

Although for $N=3,5,7$
we have not been able to compute $F^{(r,s)}(\tau, z)$
directly, a set of $F^{(r,s)}(\tau, z)$ satisfying the requirements
given above are as follows. Let us define
\be{efirst}
A(\tau, z) =  \left[ {\vartheta_2(\tau,z)^2
\over \vartheta_2(\tau,0)^2} +
{\vartheta_3(\tau,z)^2\over \vartheta_3(\tau,0)^2}
+ {\vartheta_4(\tau,z)^2\over \vartheta_4(\tau,0)^2}\right]\, ,
\ee
\be{second}
B(\tau, z) = \eta(\tau)^{-6} \vartheta_1(\tau, z)^2\, ,
\ee
and
\be{third}
E_N(\tau) = {12 i\over \pi(N-1)} \, \p_\tau \left[ \ln\eta(\tau)
-\ln\eta(N\tau)\right]= 1 + {24\over N-1} \, \sum_{n_1,n_2\ge 1\atop
n_1 \ne 0 \,  mod \, N} n_1 e^{2\pi i n_1 n_2 \tau}\, .
\ee
Then under an $SL(2,\ZZZ)$ transformation 
$A(\tau,z)$ transforms as a 
weak Jacobi form of weight 0 and index 1 and 
$B(\tau,z)$ transforms as a weak
Jacobi form of weight $-2$ and index 1. Furthermore 
\be{fourth}
E_N(\tau+1) = E_N(\tau), \qquad E_N(-1/\tau) = -\tau^2 {1\over N}
E_N(\tau/N)\, .
\ee 
{}From this it follows that $E_N(\tau)$ is a modular form of weight
2 of the group $\Gamma_0(N)$ and hence also of $\Gamma(N)$\cite{schoen}.
Using these properties one can show that the following choice of
$F^{r,s}(\tau, z)$ satisfy all the requirements described above:
\bea{fifth}
F^{(0,0)}(\tau, z) &=& {8\over N} A(\tau, z)\, ,
\nonumber \\
F^{(0,s)}(\tau, z) &=& {8\over N(N+1)} \, A(\tau, z) -{2\over N+1}
\, B(\tau, z) \, E_N(\tau) \qquad \hbox{for $1\le s\le (N-1)$}
\, , \nonumber \\
F^{(r,rk)}(\tau, z) &=& {8\over N(N+1)} \, A(\tau, z)
+ {2\over N(N+1)} \, E_N\left({\tau+k\over N}\right)\, B(\tau, z)\, ,
\nonumber \\
&& \qquad \qquad \qquad \qquad 
\qquad \hbox{for $1\le r \le (N-1)$, $0\le k\le (N-1)$}
\, . \nonumber \\
\eea

The rest of the analysis now proceeds exactly as in the $N=2$ case.  We 
arrive at an analog of eq.(\ref{bint1}) for $\II_{r,s,l}$: 
\bea{bint1newer}
\II_{r,s,l} &=& \int_\FF
\frac{d^2\tau}{\tau_2} \sum_{m_1, m_2, n_2\in \zzz,
n_1\in \zzz+{r\over N}, b\in 2\zzz + l}
\exp\left[ 2\pi i \tau( m_1 n_1 + m_2 n_2 +\frac{b^2}{4} ) \right]
\times \cr
&\;& \;\;\; \exp \left(\frac{-\pi \tau_2}{Y} \left|
n_2 ( TU -V^2) + bV + n_1 T -Um_1 + m_2 \right|^2 \right)\,
e^{2\pi i m_1 s / N} \, h^{(r,s)}_l(\tau)\, , \nonumber \\
&& \qquad \qquad \qquad \qquad
\qquad 0\le r,s\le (N-1)\, . \nonumber \\
\eea
This can then be Poisson resummed and analyzed using the techniques
described in appendix \ref{sb} and be split into holomorphic and
anti-holomorphic parts to extract the expression for $\wt\Phi_k$. On
the other hand if we want information about $\Phi_k$ we need to use
the operation eq.(\ref{ei2.6}) to consider a new integral
\bea{bint1newest}
\II_{r,s,l}' &=& \int_\FF
\frac{d^2\tau}{\tau_2} \sum_{m_1, n_1, n_2\in \zzz,
m_2\in \zzz-{r\over N}, b\in 2\zzz + l}
\exp\left[ 2\pi i \tau( m_1 n_1 + m_2 n_2 +\frac{b^2}{4} ) \right]
\times \cr
&\;& \;\;\; \exp \left(\frac{-\pi \tau_2}{Y} \left|
n_2 ( TU -V^2) + bV + n_1 T -Um_1 + m_2 \right|^2 \right)\,
e^{-2\pi i n_2 s / N} \, h^{(r,s)}_l(\tau)\, , \nonumber \\
&& \qquad \qquad \qquad \qquad
\qquad 0\le r,s\le (N-1)\, . \nonumber \\
\eea
In this case Poisson resummation over $m_1$, $m_2$ will give rise to
an additional factor of $\exp(2\pi i k_2 r / N)$ and the final sum
will be over integer values of $n_1$, $n_2$, $k_1$, $k_2$. This can
again be analyzed using the techniques described in appendix \ref{sb}.

We shall not give the details of the analysis but write down the final 
expression. The expressions for $\Phi_k$ and $\wt\Phi_k$ obtained this way 
are:
\bea{defphik}
\Phi_{k}(U,T,V) &=& -\exp\left\{2\pi i \left(
T+ U + V\right) \right\} \nonumber \\
&& \prod_{r,s=0}^{N-1}\,  \prod_{(k',l,b)\in \zzz\atop
(k', l, b)>0}
\Big\{ 1 - e^{2\pi i r / N} \, \exp( 2\pi i ( k' T + l U + b V) 
\Big\}^{{1\over 2}c^{(r,s)}(4k'l - b^2)
}  \nonumber \\
&& \prod_{r,s=0}^{N-1}\,  \prod_{(k',l,b)\in \zzz\atop
(k', l, b)>0}
\Big\{ 1 - e^{-2\pi i r / N} \, \exp( 2\pi i ( k' T + l U + b V) 
\Big\}^{{1\over 2}c^{(r,s)}(4k'l - b^2)
}  \nonumber \\
\eea
\bea{defpiknew}
\wt\Phi_{k}(U,T,V) &=& -(i\sqrt N)^{-k-2}\, 
\, \exp\left(2\pi i \left( {1\over N}\,
T+ U + V\right) \right) \nonumber \\
&&
\prod_{r=0}^{N-1}
\prod_{l,b\in \zzz, k'\in \zzz+{r\over N}\atop
k',l,b>0}\Bigg\{ 
 1 - \exp( 2\pi i( k'T + l U + bV)) \Bigg\}^{{1\over 2}\sum_{s=0}^{N-1}
e^{-2\pi ils/N}\, c^{(r,s)}(4lk' -b^2)
  }  \nonumber \\
  && \prod_{r=0}^{N-1}
\prod_{l,b\in \zzz, k'\in \zzz-{r\over N}\atop
k',l,b>0}\Bigg\{ 
 1 - \exp( 2\pi i( k'T + l U + bV)) \Bigg\}^{{1\over 2}\sum_{s=0}^{N-1}
e^{2\pi ils/N}\, c^{(r,s)}(4lk' -b^2)
  }  \nonumber \\
   \, .
\eea
$\Phi_k$ has been normalized so that the coefficient of the $\exp(2\pi
i ( T+ U + V) )$ is $-1$.  $\wt\Phi_k$ is normalized so that the
coefficient of the $\exp(2\pi i ( {1\over N}\, T+ U + V) )$ term is
$-(i\sqrt N)^{-k-2}$. These conventions agree with the one used in
\cite{0510147}.

The weight $k$ of the  modular form, determined by examining the term 
proportional to $\ln\det{\rm Im}\Omega$ in the final expression for the 
integral, is given by
\be{ekfin}
k = {1\over 2} \sum_{s=0}^{N-1} c^{(0,s)}(0) = {24\over N+1} -2\, ,
\ee
where we have used eq.(\ref{ecrsv}).
This agrees with (\ref{edefk}).
Furthermore, using eqs.(\ref{ecrsv}), (\ref{esz2}) and (\ref{esz6})
we can study the $V\to 0$ limits of $\Phi_k$ and $\wt\Phi_k$.
We get
\be{esz10}
\Phi_{k}(U,T,V) \simeq 4\pi^2 V^2 (\eta(T)\eta(NT))^{k+2}
(\eta(U)\eta(NU))^{k+2}\, ,
\ee
and
\be{esz11}
\wt\Phi_k(U,T,V) \simeq (i\sqrt N)^{-k-2}\, 4\pi^2 V^2 \, 
 (\eta(T)\eta(T/N))^{k+2}
(\eta(U)\eta(NU))^{k+2}\, ,
\ee
in agreement with \cite{0510147}.

Another important consistency check for eqs.(\ref{defphik}), 
(\ref{defpiknew})
comes from looking at the coefficient of the 
terms
involving a single power of $e^{2\pi i U}$ and all powers of $T$
and $V$. For $\Phi_k$ this is given by
\be{effe}
e^{2\pi i U}\, \eta(T)^{k-4} \, \eta(NT)^{k+2} \, \vartheta_1(T,V)^2\, ,
\ee
and for $\wt\Phi_k$ this is given by
\be{efff}
(i\sqrt N)^{-k-2} \, e^{2\pi i U}\, \eta(T)^{k-4} \,
\eta(T/N)^{k+2}
\, \vartheta_1(T,V)^2\, .
\ee
These agree with the corresponding expressions found in
\cite{0510147}.

We have also compared a few terms in the expansion of $\Phi_k$
given in (\ref{defphik}) with the one given in \cite{0510147}. The
results are given in appendix \ref{sc}.

\bigskip

\acknowledgments 
We would like to thank A.~Dabholkar, D.~Gaiotto, E.~Gava, 
R.~Gopakumar, S.~Gun, K.~S.~Narain, B.~Ramakrishnan and D.~Suryaramana for 
useful discussions.
D.~J. would like to acknowledge the hospitality of 
TIFR and APCTP where part of this work was done.
 
\appendix

\section{Calculation of the Elliptic Genus} \label{sa}

In this appendix we shall calculate
\bea{ef6app}
&& F_{m_1, m_2, n_1, n_2}(\tau,z) \nonumber \\
&=&  Tr_{RR;m_1, m_2, n_1, n_2}
\left( (-1)^{(F_{K3} +
F_{T^2})}
(-1)^{(\bar F_{K3}+\bar F_{T^2})} F_{T^2} \bar F_{T^2}
e^{2\pi i z F_{K3}} \, q^{L_0'} \bar q^{\bar L_0'}
\right)\, ,
\eea
in the superconformal field theory with target space $(K3\times 
T^2)/\ZZZ_2$. For this we shall use an orbifold description of $K3$.
We consider a  superconformal $\sigma$-model with target space
$T^2\times T^4$ with $y^1, y^2$ denoting the $T^2$ coordinates and
$y^3,y^4, y^5, y^6$ denoting the $T^4$ coordinates, and mod out the
theory by a
$\ZZZ_2\times \ZZZ_2$ symmetry generated by elements 
$g$ and $g'$. The action of
$g$ and $g'$ are given by:
\bea{ed1}
g &:& \quad
(y^1, y^2, y^3, y^4, y^5, y^6) \to (y^1, y^2, -y^3, -y^4, -y^5, -y^6)
\nonumber \\
g' &:& \quad (y^1, y^2, y^3, y^4, y^5, y^6) \to (y^1+\pi, y^2, y^3+\pi,
y^4,
y^5,
y^6) \, .
\eea
Orbifolding by $g$ produces a $K3\times T^2$ manifold. Further
orbifolding by $g'$ produces $(K3\times T^2)/\ZZZ_2$ where the
$\ZZZ_2$ generator involves a shift along $T^2$ and a $\ZZZ_2$
involution in $K3$ that preserves the (4,4) superconformal symmetry of
the corresponding world-sheet theory.  We denote by $F_{T^4}$ and
$F_{T^2}$ holomorphic parts of the world-sheet fermion number
associated with the $T^4$ and the $T^2$ parts and by $\bar F_{T^4}$
and $\bar F_{T^2}$ the anti-holomorphic parts of the world-sheet
fermion number associated with the $T^4$ and the $T^2$ parts.  We
shall be considering an arbitrary $T^2$ parametrized by the K\"{a}hler
modulus $T$ and complex structure modulus $U$, and arbitrary Wilson
lines $A_1$, $A_2$ corresponding to deforming the world-sheet theory
by the marginal operator \be{emargt} \sum_{i=1}^2 A_i \, \int d^2 z \,
\bar \p Y^i J_{T^4}\, , \ee where $J_{T^4}$ is the U(1) current
corresponding to the charge $F_{T^4}$.  We shall denote by $V$ the
complex combination $A_2-iA_1$.

We now define
\bea{ed2}
&& F_{m_1, m_2, n_1, n_2}(a,b;c,d;\tau, z) \nonumber \\
&=& Tr^{T^4\times T^2}_{m_1, 
m_2, n_1, n_2; RR; g^a, g^{\prime b}} 
\left( (-1)^{(F_{T^4} + 
F_{T^2})} 
(-1)^{(\bar F_{T^4}+\bar F_{T^2})} F_{T^2} \bar F_{T^2} e^{2\pi i z 
F_{T^4}} q^{L_0}
\bar q^{\bar L_0} g^c g^{\prime d} \right) \, ,\nonumber \\
\eea
where $L_0'$, $\bar L_0'$ have been defined in eqs.(\ref{e7}),
(\ref{ef4}).  Here $a,b,c,d$ take values 0 or 1.  $Tr^{T^4\times
  T^2}_{m_1, m_2, n_1, n_2; RR; g^a, g^{\prime b}}$ denotes trace in
the original CFT associated with the $T^2\times T^4$ target space over
RR sector states twisted by $g^a g^{\prime b}$ and carrying $(m_1,
m_2)$ units of momentum and $(n^1,n^2)$ units of winding along $(y^1,
y^2)$.  The quantity $F_{m_1, m_2, n_1, n_2}(\tau, z)$ is then given
by
\be{ed3}
F_{m_1, m_2, n_1, n_2}(\tau, z) = {1\over 4}
\sum_{a,b,c,d=0}^1  F_{m_1, m_2, n_1, n_2}(a,b;c,d;\tau, z)\, .
\ee

We shall now calculate $F_{m_1, m_2, n_1, n_2}(a,b;c,d;\tau, z)$.
First we note that
\be{ed4}
F_{m_1, m_2, n_1, n_2}(0,0;0,d;\tau, z) = 0\,  \quad \hbox{for
$d=0,1$}
\ee
due to the fermion zero modes associated with the $3,4,5,6$ directions.

Next we have
\bea{ed5}
F_{m_1, m_2, n_1, n_2}(0,0;1,d;\tau, z) &=& (-1)^{m_1 d} \, 
4\, (1+e^{2\pi i z}) (1 + e^{-2\pi i z}) \nonumber \\
&& {\prod_{n=1}^\infty (1 + q^n e^{2\pi i z})^2 (1 +  q^n e^{-2\pi i z})^2
\over \prod_{n=1}^\infty (1 + q^n)^4} \nonumber \\
&=& (-1)^{m_1 d} \, 16\,  
{\vartheta_2(\tau,z)^2\over \vartheta_2(\tau,0)^2}\, .
\eea
In the first line the factor of 4 comes from the anti-holomorphic
fermion zero modes associated with the 3,4,5,6 directions and the
factor of $(1+e^{2\pi i z}) (1 + e^{-2\pi i z})$ comes from the
holomorphic fermion zero-modes. In the second line the numerator comes
from the holomorphic non-zero mode fermionic oscillators associated
with the 3,4,5,6 directions and the denominator comes from the
holomorphic non-zero mode bosonic oscillators associated with the same
directions. The contribution from the bosonic and fermionic
oscillators associated with the 1 and 2 directions cancel.  Also the
contributions from all the non-zero mode fermion and bosonic
oscillators in the anti-holomorphic sector always cancel.  In arriving
at (\ref{ed5}) we have used that the action of $g'$ on the state
carrying $m_1$ units of momentum along $y^1$ gives a factor of
$(-1)^{m_1}$ and the action of $g$ changes the signs of the fermionic
and the bosonic oscillators associated with $T^4$.  Also since the
action of $g$ reverses the direction of momentum along the 3,4,5,6
directions, only states carrying zero momentum along $T^4$ contributes
to the trace and hence the result is independent of the moduli of
$T^4$. This will be a generic feature of all the terms; either they
will vanish due to fermion zero modes or only the zero momentum mode
will contribute due to either a $g$ insertion or a twist under $g$.

Let us now turn to the twisted sector states. First note that there
are 16 twisted sector states under $g$, located as $y^m=0, \pi$ for
$m=3,4,5,6$. $g'$ (and also $gg'$) exchanges these states pairwise.
Thus the action of $g'$ and $gg'$ on these states is off-diagonal and
hence the trace of $g'$ and $gg'$ over these states vanish. This gives
\be{ed6}
F_{m_1, m_2, n_1, n_2}(1,0;c,1;\tau, z) = 0\,  \quad 
\hbox{for $c=0,1$}\, .
\ee
On the other hand we have
\bea{ed7}
F_{m_1, m_2, n_1, n_2}(1,0;c,0;\tau, z) &=& 
16 {\prod_{n=0}^\infty (1 -
q^{n+{1\over 2}} 
e^{2\pi i z + i\pi c})^2 (1 -  q^{n+{1\over 2}} e^{-2\pi i z
+ i\pi c})^2
\over \prod_{n=0}^\infty (1 - e^{i\pi c}\, 
q^{n+{1\over 2}} )^4} \nonumber \cr
\\
&=& \cases{\, 16\,
{\vartheta_4(\tau,z)^2 / \vartheta_4(\tau,0)^2} \quad \hbox{for $c=0$}
\cr
\, 16\,
{\vartheta_3(\tau,z)^2 / \vartheta_3(\tau,0)^2} \quad \hbox{for $c=1$}}
\, .
\eea
The factor of 16 is due to the existence of 16 twisted sector states.

Next we consider sectors twisted by $g'$. In this case the winding
number $n_1$ along $y^1$ must be half integer and similarly the
winding number along $y^3$ must also be half integer. Since the $g'$
twist just involves a shift and does not affect the world-sheet
fermions, the fermion zero modes associated with the 3-6 directions
make the contribution vanish unless the $g$ projection is inserted
into the trace. This gives:
\be{ed8}
F_{m_1, m_2, n_1, n_2}(0,1;0,d;\tau, z) = 0\,  \quad 
\hbox{for $d=0,1$} \, .
\ee
On the other hand the action of $g$ as well as of $gg'$ reverses
the sign of the winding number along $y^3$ and hence these elements are 
off-diagonal in the sector twisted by $g'$. This gives
\be{ed9}
F_{m_1, m_2, n_1, n_2}(0,1;1,d;\tau, z) =0\,  \quad 
\hbox{for $d=0,1$} \, .
\ee

Finally let us turn to the sector twisted under $gg'$. Action of $gg'$
on $y^3,y^4,y^5,y^6$ gives fixed points at $y^3=\pi/2, 3\pi/2$,
$y^m=0, \pi$ for $m=4,5,6$. Although this are not real fixed points
due to the shift action $y^2\to y^2+\pi$, we can label the 16 twisted
sectors by these would be fixed points. Both $g$ and $g'$ exchange
these fixed points pairwise and hence are represented by off-diagonal
matrices.  This gives
\bea{ed10}
F_{m_1, m_2, n_1, n_2}(1,1;1,0;\tau, z) &=& 0 \, , \nonumber \\
F_{m_1, m_2, n_1, n_2}(1,1;0,1;\tau, z) &=& 0 \, .
\eea
On the other hand both the identity element and $gg'$ leave the fixed 
points invariant and give non-zero answers. We have
\bea{ed11}
F_{m_1, m_2, n_1, n_2}(1,1;0,0;\tau, z) &=&  
16\, {\prod_{n=0}^\infty (1 - q^{n+{1\over 2}} e^{2\pi i z})^2 (1 -  
q^{n+{1\over 2}} e^{-2\pi i z})^2
\over \prod_{n=0}^\infty (1 - q^{n+{1\over 2}})^4} \nonumber \\
&=&  16\,  
{\vartheta_4(\tau,z)^2\over \vartheta_4(\tau,0)^2}\, ,
\eea
and
\bea{ed12}
F_{m_1, m_2, n_1, n_2}(1,1;1,1;\tau, z) &=&
16\, (-1)^{m_1}
\, {\prod_{n=0}^\infty (1 - q^{n+{1\over 2}} e^{2\pi i 
z+i\pi})^2 (1 -
q^{n+{1\over 2}} e^{-2\pi i z + i\pi})^2
\over \prod_{n=0}^\infty (1 - e^{i\pi} q^{n+{1\over 2}})^4} \nonumber \\
&=&  16\, (-1)^{m_1} \,
{\vartheta_3(\tau,z)^2\over \vartheta_3(\tau,0)^2}\, .
\eea

Using eqs.(\ref{ed2})-(\ref{ed12}) we now get
\bea{ee1}
F_{m_1, m_2, n_1, n_2}(\tau, z) &=& 4\left[ {\vartheta_2(\tau,z)^2
\over \vartheta_2(\tau,0)^2} + 
{\vartheta_3(\tau,z)^2\over \vartheta_3(\tau,0)^2} 
+ {\vartheta_4(\tau,z)^2\over \vartheta_4(\tau,0)^2}\right] \nonumber \\
&&
+ 4 (-1)^{m_1} {\vartheta_2(\tau,z)^2\over \vartheta_2(\tau,0)^2}
\qquad\qquad\qquad \, \, \quad 
\hbox{for $n_1\in \ZZZ$} \nonumber \\
&=& 4 {\vartheta_4(\tau,z)^2\over \vartheta_4(\tau,0)^2}
+ 4\, (-1)^{m_1} \, {\vartheta_3(\tau,z)^2\over \vartheta_3(\tau,0)^2}
\qquad \hbox{for $n_1\in \ZZZ+{1\over 2}$}
\eea

\section{Evaluation of the Integral} \label{sb}

In this appendix we shall evaluate the integral (\ref{ezz5})
\bea{ezz5app}
\II &=& \sum'_A 
 \int_{\FF_A}
\frac{d^2\tau}{\tau_2^2} \,
\frac{Y}{U_2}
\, \exp\Bigg(  - \frac{\pi Y}{U_2^2 \tau_2}|{\cal A}|^2
- 2\pi i T {\rm det}\,  A \nonumber \\
&& - \frac{\pi n_2}{U_2} ( V^2
\tilde{\cal A} -
\bar V^2 {\cal A}) + \frac{2 \pi i V_2^2}{U_2^2} ( n_1 + n_2 \bar U) {\cal
A} \Bigg) \, F^{(r,s)}\left(\tau, -i { V\tilde {\cal A} -
\bar V
{\cal A} \over
2\, U_2}\right) \, .\nonumber \\
\eea
The sum over $A$ runs over all integer valued $2\times 2$ matrices of
the form (\ref{fing1}) which are not related to each other by an
$SL(2,\ZZZ)$ transformation acting from the right.  $\FF_A$ is the
union of images of the fundamental region $\FF$ under $SL(2,\ZZZ)$
transformations which act non-trivially on $A$. $\AAA$, $\tilde\AAA$
are defined in (\ref{fing2}) and $(r,s)=(2n_1, 2k_1)$ mod 2.

In carrying out the integral we need to introduce some regularization
and subtraction scheme. Following \cite{dixon} we regularize possible
divergences in the integral by including a factor of $(1 -
\exp(-\Lambda/\tau_2))$ in the integrand. For $\tau_2<<\Lambda$ this
factor is close to unity, but for $\tau_2>>\Lambda$ it is close to
zero. We also add to the integral a term
\be{esubtract}
- \left( c^{(0,0)}(0)+c^{(0,1)}(0)\right) \, \int_\FF {d^2\tau
\over \tau_2^2} \, (1 - \exp(-\Lambda/\tau_2))\, .
\ee
As we shall see, this is necessary for getting a finite $\Lambda\to\infty$
limit.

Following the same procedure as in \cite{dixon} we split the 
integration into the three orbits.
 
\noindent
{\bf{1. Contribution $\II_1$ from the zero orbit}}
\\
For $A=0$ we have $(r,s)=(0,0)$ and $\FF_A=\FF$, --
the fundamental region of $SL(2,\ZZZ)$. 
The integral (\ref{ezz5})
reduces to 
\be{bi1}
{\cal I}_1 = \frac{Y}{U_2} \int_\FF \frac{d^2\tau}{\tau_2^2} 
F^{(0,0)}(\tau, 0)
= \frac{Y}{U_2}\, \frac{\pi}{3} \, 12\, ,
\ee
using the expression for $F^{(0,0)}(\tau, z)$ given in (\ref{ex4}).
 
\noindent
{\bf{2. Contribution $\II_2$ from the non-degenerate orbit}}
\\
Here we consider $A$ to be
\be{nonde}
A = \left(
\begin{array}{cc}
k & j \\
0 & p 
\end{array}
\right), \quad 2k-1\geq 2j \geq 0, \quad p\neq 0, \quad k,j\in {1\over 2}
\ZZZ, \quad
p\in  \ZZZ \, .
\ee
In this case  the region $\FF_A$ corresponds to
two copies of the upper-half plane (coming from $A$ and
$-A$) and
 the indices $(r,s)$ in (\ref{ezz5app}) are given by
\be{ezz6}
(r,s)=(2k\,  \hbox{mod}\, 2, 2j \,  \hbox{mod}\, 2)  \, .
\ee
Note that for the above form of $A$, 
\be{edet}
\det\, A = kp, \quad {\cal A} = k\tau + j + pU,
\quad \tilde{\cal A} = k\tau + j
+ p \bar U\, .
\ee

Let us first consider the case $k\in\ZZZ$, $j\in\ZZZ$.  In this case
$j$ runs from 0 to $k-1$ in steps of 1.  The relevant $F^{(r,s)}$ is
$F^{(0,0)}$.  In order to carry out the integral we replace
$F^{(0,0)}(\tau,z)$ in (\ref{ezz5app}) by its Fourier expansion
(\ref{efourier}).  If we now change the integration variable from
$\tau_1$ to
\be{edeftaup}
\tau_1' = k\tau_1 + j + pU_1\, ,
\ee
then $\AAA$, $\tilde\AAA$ and hence also the exponential factor in
(\ref{ezz5app}), expressed as a function of $\tau_1'$ and $\tau_2$,
will have no $j$ dependence. The only $j$ dependence comes from the
term
\be{jdep}
\exp(2\pi i n \tau_1)=\exp\left( 2\pi i n 
\frac{1}{k}(\tau_1' - j- pU_1) \right)
\ee
which arises from the factor $c^{(0,0)}(4n-b^2) \exp(2\pi i \tau n)$
in the expansion (\ref{efourier}) of $F^{(0,0)}(\tau, z))$.  Since in
this case $n$ is an integer, the summation over $j$ from $0$ to $k -1$
in steps of ${1}$ imposes the condition $n = n'k$ where $n'$ is an
integer. Furthermore since $n\ge 0$ and $k>0$, we have $n'\geq 0$. The
summation over $j$ also produces a factor of $k$ which cancels the
$1/k$ factor arising due to the change of variables from $\tau_1$ to
$\tau_1'$ in the measure.

Using (\ref{edet})-(\ref{jdep}) we see that the integration over
$\tau_1'$ in (\ref{ezz5app}) is just a Gaussian integration.  The
result of carrying out this integral is
\bea{inti2}
{\cal I}_{2;k,j\in\zzz} 
&=& \sum_{n',k\in  \zzz, \, b,p\in \zzz
\atop n'\ge 0, k>0, p\ne 0}
\sqrt{Y}\int_0^\infty \frac{d\tau_2}{\tau_2^{3/2}}
\, \exp( {\cal F} ) \, c^{(0,0)}( 4n'k - b^2)
\nonumber \\
{\cal F} &\equiv& -2\pi \tau_2 n'k - 
\frac{\pi Y}{U_2^2 \tau_2}( k\tau_2 + p U_2)^2 -2\pi i T kp
-2\pi i p n'U_1
\nonumber \\
&+&  \frac{\pi b}{U_2} ( -2V_2 k\tau_2 - 2ip U_2 V_1) 
\cr
&-& \frac{2\pi V_2^2 }{U_2^2} ( k^2 \tau_2 + kp U_2) 
- \frac{\pi B^2 U_2^2 \tau_2 }{Y} \nonumber \\
B&\equiv& n' + \frac{bV_2}{U_2} + \frac{V_2^2}{U_2^2} k
\eea
The $\tau_2$ integral is of the Bessel form and can be performed
using
\be{ebessel}
\int_0^\infty {du\over u^{3/2}} \, e^{-au - bu^{-1}} = 
e^{-2\sqrt{ab}}
\sqrt{\pi\over b}\, .
\ee
This gives
\bea{finint2}
{\cal I}_{2;k,j\in\zzz} 
&=& \sum_{n',k,b\in \zzz, \, p\in  \zzz
\atop n'\ge 0, k>0, p\ne 0} \frac {1}{|p|}
 c^{(0,0)}( 4n' k - b^2) \exp\left\{  
  -2\pi i T kp - 2\pi k|p| T_2 - 2\pi kp T_2 \right.
\cr
& & - 2\pi i p n' U_1 - 2\pi |p| U_2 n' 
  \left. -2\pi i bp V_1 - 2\pi |p| b V_2 \right\}
\cr \cr
&=& -\ln \prod_{n',k,b\in \zzz\atop
 n'\geq 0, k>0}\Bigg\{  \Big| 
 1 - \exp( 2\pi i( kT + n' U + bV)) \Big|^{2 c^{(0,0)}(4n'k -b^2)
  }
 \Bigg\} \nonumber \\
\eea

Next we consider the contribution from the $k\in \ZZZ$, $j\in
\ZZZ+{1\over 2}$ terms. In this case $j$ takes values from ${1\over
  2}$ to $k -{1\over 2}$ in steps of 1 and $(r,s)=(0,1)$.  The
analysis proceeds as in the previous case, the only difference being
that the sum over $j$ of (\ref{jdep}) gives an additional factor of
$(-1)^{n'}$ besides forcing the condition $n=n'k$ with $n'\in\ZZZ$.
The analog of eq.(\ref{finint2}) is then
\be{eg1new}
{\cal I}_{2;k\in\zzz,j\in\zzz+{1\over 2}}
= -\ln \prod_{n',k,b\in \zzz\atop
n'\geq 0, k>0}\Bigg\{  \Big| 
1 - \exp( 2\pi i( kT + n' U + bV)) \Big|^{2 (-1)^{n'} 
c^{(0,1)}(4n'k -b^2)
} \Bigg\}
\ee

Finally let us consider the case $k\in\ZZZ+{1\over 2}$. In this case
instead of letting $j$ run from 0 to $k-{1\over 2}$ in steps of
${1\over 2}$ we can let it run from 0 to $(2k-1)$ in steps of 1 by
means of a further SL(2,$\ZZZ$) duality transformation.  For each of
these terms the relevant $(r,s)$ are $(1,0)$.  Proceeding as in the
$k,j\in\ZZZ$ case we now see that the sum over $j$ in (\ref{jdep})
forces the condition $n=4n'k$ with $n'\in\ZZZ$ and when this
condition is satisfied we get a factor of $2k$.\footnote{Note that
  in this case $n$ is either an integer or a half integer, but the
  sum over $j$ still forces $n$ to be an integer multiple of $k$
  since the sum runs over $2k$ values instead of $k$ values.}  The
rest of the analysis proceeds as in the previous case and we obtain
\be{eg2new}
{\cal I}_{2;k\in\zzz+{1\over 2}}
= -2\ln \prod_{n',b\in \zzz, k\in \zzz+{1\over 2}\atop
 n'\geq 0, k>0}\Bigg\{  \Big| 
 1 - \exp\{ 2\pi i( kT + n' U + bV)\} \Big|^{2 
 c^{(1,0)}(4n'k -b^2)
  }\Bigg\}
  \ee
  Thus the net contribution to the integral from the
  non-degenerate orbits take the form
  \bea{eg3}
  \II_2 &=& -\ln \Bigg[
  \prod_{n',k,b\in \zzz\atop
 n'\geq 0, k>0}\Bigg\{  \Big| 
 1 - \exp( 2\pi i( kT + n' U + bV)) \Big|^{2 c^{(0,0)}(4n'k -b^2)
 +2 (-1)^{n'}c^{(0,1)}(4n'k -b^2)  } \nonumber \\
 &&
 \prod_{n',b\in \zzz, k\in \zzz+{1\over 2}\atop
 n'\geq 0, k>0}\Bigg\{  \Big| 
 1 - \exp( 2\pi i( kT + n' U + bV)) \Big|^{4 
 c^{(1,0)}(4n'k -b^2)
  } \Bigg\}\Bigg]
\eea

\noindent
{\bf{3. Contribution $\II_3$ from the degenerate orbit} }
\\
Here we consider $A$ to be of the form
\be{dega}
A = \left(
\begin{array}{cc}
0 & j \\
0 & p 
\end{array}
\right), \quad (j, p) \neq (0,0), \quad j\in {1\over 2}
\ZZZ, \quad
p\in  \ZZZ\, .
\ee
In this case the integration region $\FF_A$ corresponds to
the strip 
\be{estrip}
 -1/2 \leq\tau_1 \leq 1/2, \qquad
\tau_2\geq 0\, . 
\ee
Also we have
\be{es1}
(r,s)=(0,0) \quad \hbox{for $j\in \ZZZ$}, \qquad 
(r,s)=(0,1) \quad \hbox{for $j\in \ZZZ+{1\over 2}$}\, .
\ee
For $A$ given in (\ref{dega})
\be{es2}
\AAA=j+pU, \qquad \tilde \AAA=j + p\bar U\, , \qquad \det A=0\, ,
\ee
are independent of $\tau$. Thus the exponential factor in (\ref{ezz5})
is independent of $\tau_1$ and the only dependence on $\tau_1$ of the
integrand comes from the $\exp(2\pi i \tau {n} )$ term in the
expansion of $F^{(r,s)}(\tau,z)$.  The $\tau_1$ integration now forces
$n$ to vanish and the coefficients $c^{(r,s)}(4n-b^2)$ multiplying the
integrand reduces to $c^{(r,s)}(-b^2)$.  It follows from the
definition of $c^{(r,s)}(m)$ that these coefficients are non-zero only
for $b =0$ and $b=\pm 1$.

We first consider the case $j\in \ZZZ$.  We begin with the
contribution from the $n =0, b=0$ term and proceed as in \cite{dixon}.
We multiply the integrand with the regulating factor $(1-\exp
(-\Lambda/\tau_2))$, then integrate over $\tau_2$ and finally perform
the sum over $j$ and $p$.  Integrating over $\tau_2$ we obtain
\bea{fbint3}
{\cal I}_{3, b=0;j\in \zzz} &=& 
c^{(0,0)}(0) \Bigg[\frac{U_2}{\pi}\, 
\sum_{(j, p)\neq (0,0)\atop j,p\in \zzz} \left(
\frac{1}{|j + Up|^2 } - \frac{1}{|j + Up|^2 + \Lambda U_2^2/\pi Y} \right)
\nonumber \\
&& - \int_{\cal F}  
d^2\tau  \frac{1- \exp( -\frac{\Lambda}{\tau_2}) }{\tau_2} \Bigg]\, .
   \eea
Note that we have introduced a subtraction  term
proportional to $\int_{\cal F}  
d^2\tau  \frac{1- \exp( - {\Lambda}/{\tau_2}) }{\tau_2}$ 
in eq.(\ref{fbint3}), -- this is one of the two terms appearing
in (\ref{esubtract}).
This is necessary in order to get a finite value of the integral in
the $\Lambda\to\infty$ limit.
The result of the integration in the second terms inside the
square brackets  is
$\ln \Lambda + \gamma_E + 1 + \ln ( 2/3\sqrt{3})$. 
To evaluate the summation we use\cite{9510182}
\bea{sumfo}
\sum_{j \in\zzz}  \frac
{\exp ( i \theta j ) }{ ( j + B )^2 + C^2 } 
&=& \frac{\pi}{C} \exp( - i \theta( B - i C))   
\frac{1}{ 1 - \exp ( - 2\pi i ( B-iC) ) } 
  \nonumber \\
&+& \frac{\pi}{C} \exp( - i \theta  ( B + i C) ) 
\frac{ \exp ( 2\pi i (B + iC) ) }{ 1 - \exp ( 2\pi i (B + iC) ) }
 \nonumber \\
& & \qquad \qquad \hbox{for} \quad
C> 0, \quad 0 \leq \theta \leq 2\pi \nonumber \\
\sum_{ j \in\zzz\atop j>0} \frac{ \cos \theta j }{ j^2} 
&=& 
\frac{\theta ( \theta - 2\pi ) }{4} + \frac{\pi^2}{6}\, .
\eea
We now regroup the summation in \eq{fbint3} as
$\sum_{p=0, j \neq 0} + \sum_{j =-\infty, p \neq0}^{j = +\infty} $
and use (\ref{sumfo}) at $\theta=0$ to obtain
\bea{suminfo}
\II_{3,b=0;j\in \zzz}&=& c^{(0,0)}(0) \Bigg[ \frac{\pi}{3} U_2 + 
\sum_{p>0\atop p\in \zzz} \Bigg\{
\frac{2}{p} 
\frac{e^{-2\pi i p \bar U} } { 1- e^{-2\pi i p \bar U}} 
+
 \frac{2}{p} 
\frac{e^{2\pi i p  U} } { 1- e^{2\pi i p  U}} 
\nonumber \\
&& +
 \left(
\frac{2}{p} -  \frac{2}{ \sqrt{ p^2 + \Lambda /\pi Y } } \right)
\Bigg\}
 -    \left( \ln \Lambda + \gamma_E + 1 + \ln ( 2/3\sqrt{3}) \right)
\Bigg]\, . \nonumber \\
\eea
Next we expand
\be{eexpand}
{x\over 1-x}=\sum_{l=1}^\infty x^l\, ,
\ee
for $x=e^{-2\pi i pU}$ and $x=e^{2\pi i p\bar U}$ in (\ref{suminfo})
and perform the sum over $p$ in the first two terms. Finally we use
\be{euse}
\sum_{p>0\atop p\in \zzz} \left(
\frac{2}{p} -  \frac{2}{ \sqrt{ p^2 + \Lambda /\pi Y } } \right)
=-\ln{\pi Y\over \Lambda} + 2\gamma_E - \ln 4\,  \quad \hbox{as
$\Lambda\to\infty$}\, ,
\ee
to obtain
 \be{fbint31}
{\cal I}_{3, b=0;j\in\zzz} =
  c^{(0,0)}(0)  \,
\left(\frac{\pi}{3}   U_2 - \ln Y   
+ \kappa' \right)   - \ln \prod_{l\in\zzz,l>0} \left\{
 \left| 1 - \exp ( 2\pi i l U ) \right|^{ 4  
  c^{(0,0)}(0)  }  \right\}
 \ee
where 
\be{edefkappa}
\kappa' = \gamma_E -1 - \ln ( 8\pi/3\sqrt 3) \, . 
\ee

We now evaluate the contribution of $n =0$, 
$b =\pm 1$. The corresponding coefficient is $c^{(0,0)}(-1)$.
Integrating over
$\tau_2$ we obtain
\be{sbint3}
{\cal I}_{3, b=\pm 1;j\in\zzz} 
= c^{(0,0)}(-1) \frac{U_2}{\pi} \sum_{(j, p)\neq (0,0)\atop
j,p\in\zzz} 
\frac{1}{|j+ pU|^2} 
\exp\left( \frac{2\pi i b}{U_2} ( j V_2 + p (V_2 U_1 - V_1 U_2) ) 
\right)  
\ee
We split this summation as before 
$\sum_{p =0, j \neq 0} + \sum_{p \neq 0, j}$.
We shall assume, for definiteness, that
\be{ev2}
V_2 < 0\, .
\ee 
For the $p=0$ one can apply the second formula in \eq{sumfo} to obtain
\be{pzeroca}
 4\pi \,c^{(0,0)}(-1)
 \, \left( \frac{V_2^2}{U_2} + V_2 + \frac{U_2}{6} 
\right)
\ee
Let us now turn to the contribution from the $p\ne 0$ terms. Since
(\ref{sbint3}) contains the contribution for both $b=1$ and $b =-1$,
care should be taken so that the $\theta$ in $\eq{sumfo}$ is between
$0\leq \theta \leq 2\pi$. Here $\theta = -2\pi V_2/U_2 \leq 1$.  For
the $p\neq 0$ case one splits the summation for $p>0, b =\pm 1$ and
$p<0, b= \pm 1$, then one changes $j \rightarrow -j$ or $p\rightarrow
-p$ so that one can always apply the formula in \eq{sumfo}.  Carefully
taking all these contributions into account one obtains, after using
\eq{sumfo}, the total contribution from the $p\ne 0$ terms to be
\be{allsump}
  - \ln \prod_{l\in\zzz, l>0, b =\pm 1} 
| 1- \exp ( 2\pi i ( lU + bV) |^{ 4   c^{(0,0)}(-1)
 }   - \ln | 1- \exp ( -2\pi i V)  |^ {4  
c^{(0,0)}(-1) } 
 \ee
Thus the net contribution from the $b=\pm1$, $j\in\ZZZ$ terms are
\bea{enet}
{\cal I}_{3,b=\pm 1;j\in\zzz}
&=& 4\pi 
\, c^{(0,0)}(-1) \, 
\left( \frac{V_2^2}{U_2} + V_2 + \frac{U_2}{6}  \right)
\nonumber \\
&& - \ln \prod_{l\in\zzz, l>0 \atop 
b =\pm 1} \left\{ \left| 1- \exp ( 2\pi i ( lU + bV)) \right|^{ 4 
 c^{(0,0)}(-1)   } \right\} \nonumber \\
&& 
 - \ln \left| 1- \exp ( -2\pi i   V) \right|^{ 4 
 c^{(0,0)}(-1)   }
\eea
Note that the last term in the above equation 
is singular as $V\to 0$.

Next we turn to the contribution from the $j\in\ZZZ+{1\over 2}$ terms.
In this case $(r,s)=(0,1)$. The analog of (\ref{sumfo}) is obtained by
replacing $B\to B+{1\over 2}$ in this formula and multiplying the
resulting equation by a factor of $e^{i\theta/2}$ on both sides:
\bea{newsumfo}
\sum_{j \in \zzz+{1\over 2}}  \frac
{\exp ( i \theta j ) }{ ( j + B )^2 + C^2 } 
&=& \frac{\pi}{C} \exp( - i \theta( B - i C))  
\frac{1}{ 1 + \exp ( - 2\pi i ( B-iC) ) } 
  \nonumber \\
&-& \frac{\pi}{C} \exp( - i \theta  ( B + i C) ) 
\frac{ \exp ( 2\pi i (B + iC) ) }{ 1 + \exp ( 2\pi i (B + iC) ) }
 \nonumber \\
& & \qquad \qquad \hbox{for} \quad
C> 0, \quad 0 \leq \theta \leq 2\pi  
\eea
Using this result 
we can get the analogs of (\ref{fbint31}) and
(\ref{enet}):
\be{fbint31new}
{\cal I}_{3, b=0;j\in\zzz+{1\over 2}} =
  c^{(0,1)}(0)  \,
\left( {\pi}    U_2 - \ln Y   
+ \kappa' \right)   - \ln \prod_{l\in\zzz,l>0} \left\{
 \left| 1 - \exp ( 2\pi i l U ) \right|^{ 4  (-1)^l
  c^{(0,1)}(0)  }  \right\}
 \ee
\bea{enetnew}
{\cal I}_{3,b=\pm 1;j\in\zzz+{1\over 2}}
&=&4\pi 
\, c^{(0,1)}(-1) \, 
\left(  V_2 + \frac{U_2}{2} \right) \nonumber\\
&& - \ln \prod_{l\in\zzz, l>0 \atop 
b =\pm 1} \left\{ \left| 1- \exp ( 2\pi i ( lU + bV) \right|^{ 4 
(-1)^l
 c^{(0,1)}(-1)   } \right\} \nonumber \\
 && 
 - \ln \left| 1- \exp ( -2\pi i   V) \right|^{ 4 
 c^{(0,1)}(-1)   }
 \eea

Adding all the contributions we obtain.
\bea{tdego}
{\cal I}_{3 } &=& {\cal I}_{3,b=0;j\in\zzz}
+ {\cal I}_{3,b=\pm 1;j\in\zzz } 
+{\cal I}_{3,b=0;j\in\zzz+{1\over 2}}
+ {\cal I}_{3,b=\pm 1;j\in\zzz+{1\over 2}}\nonumber \\
&=&  c^{(0,0)}(0)  \,
\left(\frac{\pi}{3}   U_2 - \ln Y  
+ \kappa' \right)  + 4\pi 
\, c^{(0,0)}(-1) \, 
\left( \frac{V_2^2}{U_2} + V_2 + \frac{U_2}{6} \right)
\nonumber \\
&& +   c^{(0,1)}(0)  \,
\left( {\pi}    U_2 - \ln Y   
+ \kappa' \right) + 
4\pi 
\, c^{(0,1)}(-1) \, 
\left(  V_2 + \frac{U_2}{2} \right)
\nonumber \\
&& - \ln \prod_{l\in\zzz,l>0} \left\{
 \left| 1 - \exp ( 2\pi i l U ) \right|^{ 4  
  c^{(0,0)}(0)  } \right\} -\ln 
  \left\{| 1- \exp ( -2\pi i V ) |^ {4   c^{(0,0)}(-1) 
  }    \right\} \nonumber \\
  &&
 - \ln \prod_{l\in\zzz, l>0 \atop 
b =\pm 1} \left\{ \left| 1- \exp ( 2\pi i ( lU + bV) \right|^{ 4 
 c^{(0,0)}(-1)   } \right\}   \nonumber \\
&&  - \ln \prod_{l\in\zzz,l>0} \left\{
 \left| 1 - \exp ( 2\pi i l U ) \right|^{ 4  (-1)^l
  c^{(0,1)}(0)  }  \right\} - \ln \left| 1- \exp ( -2\pi i   V) \right|^{ 
4 
 c^{(0,1)}(-1)   }  \nonumber \\
 && 
  - \ln \prod_{l\in\zzz, l>0 \atop 
b =\pm 1} \left\{ \left| 1- \exp ( 2\pi i ( lU + bV) \right|^{ 4 
(-1)^l
 c^{(0,1)}(-1)   } \right\} 
\eea

Combining the contribution from all the orbits and noting that
\bea{ecvalue}
&& c^{(0,0)}(0)=10, \quad
c^{(0,0)}(-1)=1,\quad  c^{(0,1)}(0)=2,  
\quad  c^{(0,1)}(-1)=1, \nonumber \\
&&c^{(1,0)}(0)=4, \quad c^{(1,0)}(-1)=0, 
\quad c^{(1,1)}(0)=4, \quad  
c^{(1,1)}(-1)=0\, ,
\eea
we can now express the full integral as
\bea{finalfulk3app-}
{\cal I} &=& {\cal I}_1 + 2 {\cal I}_2 + {\cal I}_3, \cr
&=& - 2 \ln \Bigg[
{\kappa}\,  (\det{\rm Im}\Omega)^{6} \Bigg|
\exp( 2\pi i ( {1\over 2}T+ U+ V) ) \nonumber \\
&& \prod_{(k,l.b)\in \zzz\atop
(k, l, b)>0} 
( 1 - \exp( 2\pi i ( k T + l U + b V) ) ^{c^{(0,0)}(4kl - b^2)
+ (-1)^l\, c^{(0,1)}(4kl - b^2)} \nonumber \\
&& \prod_{l,b\in \zzz, k\in \zzz+{1\over 2}\atop
 l\geq 0, k>0}\Bigg\{  \left| 
 1 - \exp( 2\pi i( kT + l U + bV)) \right|^{2
 c^{(1,0)}(4lk -b^2)
  } \Bigg\} \Bigg |^2 \Bigg] 
  \eea
  where
  \be{ekap}
 \kappa = \left ( \frac{8\pi}{ 3\sqrt 3 } e^{1-\gamma_E} \right)^{6}
\ee
and $(k, l , b)>0$ means $k> 0 , l\geq 0, b \in \ZZZ$ or $ k =0, l>0,
b \in \ZZZ$ or $k =0, l =0, b <0$.  Note that we have $2{\cal I}_2$
because of the two copies of the upper half plane.

{}From the modular transformation laws (\ref{ex2a}) and the series
expansion (\ref{efourier}) it follows that
\be{ecprop}
c^{(1,1)}(4lk - b^2) = (-1)^l \, c^{(1,0)}(4lk-b^2) \qquad
\hbox{for $k\in \ZZZ+{1\over 2}$, $l\in \ZZZ$}\, .
\ee
Using this we can reexpress (\ref{finalfulk3app-}) in a more symmetric
fashion: 
\bea{finalfulk3app}
{\cal I}  
&=& - 2 \ln \Bigg[
{\kappa}\, (\det{\rm Im}\Omega)^{6} \Bigg|
\exp\left( 2\pi i \left( {1\over 2}T+ U+ V\right) 
\right) \nonumber \\
&& \prod_{r,s=0}^1\, \prod_{(l,b)\in \zzz, k\in \zzz+{r\over 2}
\atop
(k, l, b)>0} \, \Bigg\{
( 1 - \exp( 2\pi i ( k T + l U + b V) ) ^{(-1)^{ls}\, 
c^{(r,s)}(4kl - b^2)}  \Bigg\} \Bigg |^2 \Bigg] \, . \nonumber \\
  \eea

\section{Explicit Results for $a(n,m,r)$} \label{sc}

In this appendix we present the results of explicit computation of the
coefficients $a(n,m,r)$ for $\Phi_k$. These were calculated using the
expression given in \cite{0510147} as well as the expression found in
the present paper and found to be the same.  To write the expansion of
$\Phi_k$ in a convenient way we define $t=\exp( 2\pi i T),\, u = \exp(
2\pi i U),\, v = \exp(2\pi i V)$.  Then for $N=2$
\bea{eexpansion}
\Phi_6 &=& \left[ 
( 2 -\frac{1}{v} -v ) u + 
( -4 + \frac{2}{v^2} + 2v^2) u^2 
+ ( -16 -\frac{1}{v^3} - \frac{4}{v^2} +\frac{13}{v} + 13v 
-4 v^2 -v^3) u^3 \right] t \cr
&+&
\left[ 
( -4 + \frac{2}{v^2} + 2v^2)u +
( 32 - \frac{16}{v^2} - 16v^2 ) u^2 
+ ( -72 -\frac{4}{v^4} + \frac{40}{v^2} + 40v^2 - 4 v^4) u^3
\right] t^2 \cr
&+& \left[
 ( -16 -\frac{1}{v^3} - \frac{4}{v^2} +\frac{13}{v} + 13v 
-4 v^2 -v^3) u \right. \cr 
&+& 
 ( -72 -\frac{4}{v^4} + \frac{40}{v^2} + 40v^2 - 4 v^4) u^2
\cr
&+&
\left.
\left( 336 + \frac{13}{v^5} + \frac{40}{v^4} - \frac{87}{v^3}
-\frac{64}{v^2} -\frac{70}{v} - 70v - 64v^2 - 87v^3
+ 40v^4 + 13 v^5 \right) u^3 \right]t^3 \cr
&+& \cdots
\eea
For $N=3$
\bea{ephi4}
\Phi_4 &=& \left( 
( 2 -\frac{1}{v} - v ) u + ( \frac{2}{v^2} - \frac{2}{v}
- 2v + 2 v^2 ) u^2  \right) t \cr
&+& \left(
( \frac{2}{v^2} - \frac{2}{v} - 2v + 2 v^2) u + 
( 4 - \frac{2}{v^3} -\frac{6}{v^2} +  \frac{6}{v} 
+ 6 v - 6v^2 - 2 v^3) u^2 \right) t^2 + \cdots \nonumber \\
\eea
For $N=5$
\bea{ephi2}
\Phi_2 &=& \left(
( 2 -\frac{1}{v} - v) u + ( 4 + \frac{2}{v^2} -\frac{4}{v} - 4v
+ 2 v^2 ) u^2 \right) t \cr
&+& \left(
( 4 + \frac{2}{v^2} - \frac{4}{v} - 4 v + 2v^2) u +
( 28 - \frac{4}{v^3} + \frac{10}{v^2} - \frac{20}{v} 
- 20 v + 10 v^2 - 4 v^3) u^2 \right) t^2 + \cdots \nonumber \\
\eea
For $N=7$
\bea{ephi1}
\Phi_1 &=& \left(
( 2 -\frac{1}{v} - v) u + 
( 6 + \frac{2}{v^2} - \frac{5}{v} - 5v + 2 v^2) u^2 \right)t
\cr
&+& \left(
( 6 + \frac{2}{v^2} - \frac{5}{v} - 5v + 2 v^2) u 
+ ( 52 - \frac{5}{v^3} + \frac{19}{v^2} - \frac{40}{v} -40v + 19 v^2
- 5 v^3 ) u^2 \right) t^2 + \cdots \nonumber \\
\eea

\end{document}